\documentclass[useAMS,usenatbib,usegraphicx]{mn2e}
\usepackage{url}
\usepackage{hyperref}
\usepackage{rotating}
\usepackage{subfig}

\def\aj{AJ}%
\def\apj{ApJ}%
\def\apjs{ApJS}%
\def\apss{Ap\&SS}%
\def\aap{A\&A}%
\def\aapr{A\&A~Rev.}%
\def\aaps{A\&AS}%
\def\mnras{MNRAS}%
\def\pasp{PASP}%
\def\nat{Nature}%
\def\mum{{$\rm \mu$m}}
\def\nsamp{28 }

\def\hi{H\,{\sc i}}
\def\m20{$M_{20}$}

\begin{document}

\title[Multi-Wavelengths Morphology of the THINGS Galaxies]{Quantified \hi \ Morphology I: Multi-Wavelengths Analysis of the THINGS Galaxies.}

\author[B.W. Holwerda et al.]{B. W. Holwerda$^{1,2}$\thanks{E-mail:
benne.holwerda@esa.int}, N. Pirzkal,$^{3}$ W.J.G. de Blok,$^{2}$ A. Bouchard,$^{4}$ S-L. Blyth,$^{2}$
\newauthor
K. J. van der Heyden,$^{2}$ and E. C. Elson$^{2}$\\
$^{1}$ European Space Agency, ESTEC, Keplerlaan 1, 2200 AG, Noordwijk, the Netherlands\\
$^{2}$ Astrophysics, Cosmology and Gravity Centre (ACGC), \\
Astronomy Department, University of Cape Town, Private Bag X3, 7700 Rondebosch, Republic of South Africa\\
$^{3}$ Space Telescope Science Institute, 3700 San Martin Drive, Baltimore, MD 21218, USA\\
$^{4}$ Department of Physics, Rutherford Physics Building, McGill University, 3600 University Street, Montreal, Quebec, H3A 2T8, Canada}

\date{Accepted  Received ; in original form }

\pagerange{\pageref{firstpage}--\pageref{lastpage}} \pubyear{2010}

\maketitle

\label{firstpage}

\begin{abstract}
Galaxy evolution is driven to a large extent by interactions and mergers with other galaxies and the gas in galaxies is extremely sensitive to the interactions. One method to measure such interactions uses the quantified morphology of galaxy images. Well-established parameters are Concentration, Asymmetry, Smoothness, Gini, and $M_{20}$ of a galaxy image. Thus far, the application of this technique has mostly been restricted to restframe ultra-violet and optical images. However, with the new radio observatories being commissioned (MeerKAT, ASKAP, EVLA, WSRT/APERTIF, and ultimately SKA), a new window on the neutral atomic hydrogen gas (\hi) morphology of a large numbers of galaxies will open up. The quantified morphology of gas disks of spirals can be an alternative indicator of the level and frequency of interaction. The \hi \ in galaxies is typically spatially more extended and more sensitive to low-mass or weak interactions. 

In this paper, we explore six morphological parameters calculated over the extent of the stellar (optical) disk and the extent of the gas disk for a range of wavelengths spanning UV, Optical, Near- and Far-Infrared and 21 cm (\hi) of \nsamp \ galaxies from The \hi \ Nearby Galaxy Survey (THINGS). 
Though the THINGS sample is small and contains only a single ongoing interaction, it spans both non-interacting and post-interacting galaxies with a wealth of multi-wavelength data. 
We find that the choice of area for the computation of the morphological parameters is less of an issue than the wavelength at which they are measured. 
The signal of interaction is as good in the \hi \ as in any of the other wavelengths in which morphology has been used to trace the interaction rate to date, mostly star-formation dominated ones (near- and far-ultraviolet). The Asymmetry and $M_{20}$ parameters are the ones which show the most promise as tracers of interaction in 21 cm line observations. 
\end{abstract}


\section{Introduction}
\label{s:intro}

Evolution of galaxies in the cold dark matter with a cosmological constant  Universe ($\Lambda$CDM) appears to be driven by the merger and interaction of dark matter haloes \citep[e.g., the Millennium Simulation by][]{Springel05}. Therefore, a substantial observational effort has been made to quantify the rate of mergers and interactions over time. Several methods have been developed to estimate the interaction rate: identification of physically close pairs of galaxies in redshift surveys \citep[e.g,][]{Patton00,de-Ravel09}, measures of H-alpha equivalent width, far-IR flux from (ultra) luminous infrared galaxies \citep[(U)LIRGs, see e.g.,][]{Murphy01}, from the star-burst, OH megamasers \citep[e.g.,][]{Klockner05,Darling06} and identification of galaxies with disturbed morphologies \citep[e.g.,][]{CAS,Lotz04,Lotz08b,Conselice09b}. 
This observational effort has been matched by theoretical ones to accurately map the well-understood merger trees of Dark Matter haloes onto galaxy-galaxy merger rates \citep[see the review in][and references therein]{Hopkins10}. Thusfar, theoretical models suffer from large systematics. However, ongoing efforts in both cosmological hydrodynamical simulations and semi-analytical models can be expected to match the observational accuracy soon. 

Merger rates from disturbed morphologies of galaxies have been explored extensively with quantified classification of galaxies. Based on a series of scale-invariant parameters, quantified galaxy morphology has been applied predominantly on restframe ultra-violet images of galaxies \citep[e.g.,][]{Taylor-Mager07}. 
The advantages for optical or ultraviolet are that interacting galaxies are star-forming and hence bright in ultra-violet and blue side of the optical. Their morphology shows enhanced surface brightness and clear signs of disturbance. The high surface brightness ensures more complete samples for a given observation. Observationally this approach is also attractive as it has the advantage of similar spatial resolution at high- and low-redshift, using the Hubble Space Telescope and Galaxy Evolution Explorer (GALEX) or Sloan Digital Sky Survey (SDSS) respectively. Disadvantages of this method are the time-lag for the interaction to trigger star-formation, and modification of the morphology due to dust obscuration.

However, new windows for quantified morphology will be opening up; the far-infrared emission from star-formation and molecular gas are now resolved with Herschel and in the near future with the Atacama Large Millimeter/submillimeter Array (ALMA). The atomic gas with its many signatures of interaction (tails, bridges, beards, warps, clouds etc.) will also be much better resolved with the commissioning of the Square Kilometer Array \citep[SKA;][]{ska}, and its precursors, the South African Karoo Array Telescope \citep[MeerKAT;][]{MeerKAT,meerkat1,meerkat2}, the Australian SKA Pathfinder \citep[ASKAP;][]{askap}, and the pathfinders, the Extended Very Large Array \citep[EVLA;][]{evla}, and the APERture Tile In Focus instrument \citep[APERTIF;][]{apertif} on the Westerbork Synthesis Radio Telescope (WSRT). Spatial resolutions of these observations will start to rival those in the ultra-violet. \hi \ morphology also appears to be very sensitive to the smaller interactions with tidal features often reported to be much more visible than in any other wavelength \citep[see the ``\hi \ Rogues Gallery"\footnote{\url{http://www.nrao.edu/astrores/HIrogues/}} compilation in][]{Hibbard01}. Hence the morphology of galaxies in other wavelengths, especially \hi \ might be an equally good or surpassing indicator of tidal interaction than that of the ultra-violet and other star-formation dominated emission. Notably, the envisaged all-sky surveys with ASKAP (WALLABY\footnote{Widefield ASKAP L-band Legacy All-sky Blind surveY}), and the WSRT/APERTIF will then provide an accurate census of mergers in the local Universe. 

The aims for this paper are to explore (1) which wavelength shows the strongest signal of interaction, (2) which morphological parameters are the optimal discriminators for interaction, and (3) over which area morphological parameters need to be computed. In a companion letter \citep{Holwerda10a}, we briefly highlight how well \hi \ morphology shows the signal of interaction compared to the UV, optical and FIR. 
In further papers, we define the \hi \ parameter space to identify interacting galaxies \citep{Holwerda10c}, using a sub-sample of WHISP\footnote{The Westerbork \hi \ Spirals and irregulars Project \citep{whisp,whisp2}}, we derive a time-scale for interactions to reside in this parameter space \citep{Holwerda10d}, and we infer the first interaction fraction and rate based on the WHISP survey \citep{Holwerda10e}. The \hi \ morphological phenomena in the Virgo cluster environment (e.g., ram-pressure stripping) are explored in \cite{Holwerda10f}.

In \S \ref{s:data} we discuss our sample and data used, in section \ref{s:morph} we discuss the morphological parameters, as well as effects of uncertainty and possible biases. 
Our results are in \S \ref{s:results}, together with notes on each individual galaxy. Our discussion on the suitability of the \hi \  parameters is in \S \ref{s:disc} and our conclusions are summarized in \S \ref{s:concl}.

\section{Sample \& Data}
\label{s:data}

We use the public datasets from the \hi \ Nearby Galaxy Survey \citep[THINGS;][]{Walter08}\footnote{\url{http://www.mpia-hd.mpg.de/THINGS/}}, 
the Spitzer Infrared Nearby Galaxies Survey \citep[SINGS,][]{sings}\footnote{\url{http://sings.stsci.edu/}}, the GALEX Nearby Galaxy Atlas (NGA)\footnote{\url{http://galex.stsci.edu/GR4/}} and the Sloan Digital Sky Survey Data Release 7 (SDSS, DR7)\footnote{\url{http://cas.sdss.org/dr7/}}. Table A2 ({\em in the online version})
lists the availability of Spitzer, GALEX and SDSS data for our sample, as well as the basic data  from \cite{Walter08}.
Optical data is either from the SINGS ancillary data, Nasa Extragalactic Database\footnote{NED, \url{http://nedwww.ipac.caltech.edu/}} public data or SDSS, preferring SDSS. 
We aim to cover the sample from \cite{Trachternach08} and \cite{de-Blok08} because for these galaxies there is detailed dynamical information (rotation curves, dynamical centre, and parameterizations of non-circular motions). We require \hi \ data from the THINGS sample, Spitzer data from the SINGS sample of a program with equivalent quality data, and at least some optical data. These requirements narrow the 34 galaxies from \cite{Walter08} down to 28. 

The \hi \ column density maps are the naturally-weighted (NA) and robust-weighted (RO) total intensity maps, expressed as \hi \ column density using the formalism and beam sizes reported in \cite{Walter08}. 
We chose the naturally-weighted maps to define our contours as these are more sensitive to larger scale structures (of the order of the disk) than the robust weighted maps. The natural-weighted maps have lower spatial resolution (typically of the order of 10" resolution) than the robust-weighted ones (typically 6"). We compute morphological parameters for both \hi \ maps.
To define the area over which parameters are computed, we picked two \hi \ column density contours: 0.3 and $30 \times 10^{20}$ cm$^{-2}$. These correspond to approximately the spatial extent of the \hi \ and stellar disk respectively, but may exclude areas corresponding to "\hi \ holes".

The majority of the Spitzer IRAC and MIPS data are from the SINGS project with three additions from the Local Volume Legacy Survey \citep[LVL,][]{lvl}; NGC 5236 (M83), NGC 5457 (M101) and IC 2574. Spitzer data includes all IRAC (3.6, 4.5, 5.6 and 8.0 \mum) and MIPS (24, 70 and 160 \mum) channels. These tend to comfortably cover the stellar disk and most of the \hi \ outer contour ($3 \times 10^{19}$ cm$^{-2}$).

In the case of the SDSS optical data, we obtained the original tiles around our galaxies and combined them into larger mosaics using swarp (\url{http://astromatic.iap.fr/}). The images were sky-subtracted before combination to account for the different sky values in each scan-strip. The SDSS mosaics have distinct advantages over the NED and SINGS ancillary data: uniform depth, a well-defined set of filters and a field-of-view that covers the whole \hi \ map. 

Most of the GALEX data is from the Nearby Galaxy Atlas (NGA) supplemented with the all-sky survey in two cases: NGC 3184 and NGC 6946. NGA data generally means the galaxy is in the focus of GALEX with 4" resolution in FUV (1528 \AA) and NUV (2271 \AA). However, galaxies such as NGC 3031 and M81 Dwarf A are part of the M81/M82 group portrait and M83 is in the corner of the GALEX FOV because of a bright foreground star and these galaxies are therefore slightly out-of-focus with GALEX with resolutions closer to 6"

Table A1 ({\em online edition}) 
lists the resolutions and wavelength of all our data. GALEX resolution ranges from 4"-6" depending on position in the field, MIPS at 24 \mum \ is 6" and the RO maps are typically also 6" resolution, depending on position in the sky and axis. The MIPS 70 and 160 \mum \ and the NA maps are of poorer resolution.

\subsection{Data Preparation}
\label{ss:prep}

To start, we shift the natural weighted \hi \ map, such that the centre of the galaxy is in the centre of the image. Because the THINGS observations are pointed, the galaxy is already close to the central part of the image and the shifts are small (a few pixels). We use the central positions reported in \cite{Walter08}, who list the dynamical central positions from \cite{Trachternach08}, supplemented on occasion with Spitzer 3.6 \mum \  central positions.
Subsequently, we align all the different data using \texttt{wcsmap} and \texttt{geotran} in IRAF to this centered natural weighted (NA) \hi \ map. Our next step is to convolve the optical and 
IRAC data to 6", approximately the resolution of the majority of the rest of the data, before determining the morphological parameters.  

After alignment and smoothing, a  mask of the foreground stars is created using a Source Extractor \citep{se,seman} catalog and resulting segmentation map of the SDSS-i image, selecting small objects in the field. If the SDSS-i image is not available, we use the IRAC channel 2 (4.5 \mum). Channels 1 and 2 of the IRAC instrument trace the old red stellar component and we chose channel 2 as it does not contain the 3.1 \mum \ PAH feature so any hot ISM region that belongs to the galaxy is spuriously rejected. 
 The disadvantage of the IRAC data is that it may not cover the entire gas disk but to mask foreground stars, it is preferable to use optical or near-infrared rather than GALEX data.

\section{Morphological Parameters}
\label{s:morph}

The morphological parameters we use here have been established with repeated applications to deep Hubble Space Telescope images and reference local galaxy samples. There is the Concentration-Asymmetry-Smoothness classification scheme and the Gini and $M_{20}$ parameters. Ellipticity is sometimes added. We use the definitions shared by \cite{CAS}, \cite{Lotz04}, and \cite{Scarlata07} for our computation of the six morphological parameters: Concentration, Asymmetry, Smoothness, Gini, the moment of light ($M_{20}$) and Ellipticity.

\subsection{CAS}
\label{ss:cas}

\cite{Abraham94, Abraham96b, Abraham96a} introduced definitions of asymmetry, concentration and contrast to classify galaxies in the Hubble Deep Field North. Following the work by \cite{Bershady00} and \cite{Trujillo01a, Trujillo01}, \cite{Conselice00a} refined these parameters, culminating in \cite{CAS} which added a local volume reference in R-band \citep[the sample from ][]{Frei96}. The thus established parameter space has been used extensively on all HST wide and deep surveys; e.g, in GOODS by \cite{Bundy05} and \cite{Ravindranath06}, the HUDF by \citep{Yan05}, COSMOS by \cite{Scarlata07}, GEMS by \cite{Jogee09}, and the extended Groth strip by \cite{Trujillo07}.

\subsubsection{Concentration}
\label{sss:C}

Concentration is defined by \cite{Trujillo01b} as:
\begin{equation}
C = 5 ~ log \left( {r_{80} \over  r_{20} } \right)
\label{eq:c}
\end{equation}
\noindent with $r_{f}$ is the radius containing a percentage $f$ of the light of the galaxy, in this case 80 and 20 percent respectively. Other definitions using $r_{90}$ and $r_{50}$ are also in vogue, notably in the SDSS catalogue \citep[see the discussion in][]{Graham05}. The radii are often taken from Source Extractor \citep{se,seman} output but we computed these here without the use of the Source Extractor program. The two apertures for these radii are circular and hence this parameter is somewhat sensitive to the inclination of the spiral disk \citep[See also][and \S \ref{ss:incl}]{Scarlata07,Bendo07}. 
Concentration depends on the adopted central positionÊfor the measurement apertures.

\subsubsection{Asymmetry}
\label{sas:A}

\noindent Following earlier work by \cite{Abraham96a}, the now most commonly used definition of Asymmetry is from \cite{CAS}:
\begin{equation}
A = {\Sigma_{i,j} | I(i,j) - I_{180}(i,j) |  \over \Sigma_{i,j} | I(i,j) |  }
\label{eq:a}
\end{equation}
\noindent where $I(i,j)$ is the value of the pixel at the position $i,j$ in the image, and $I_{180}(i,j)$ is the value of the pixel in the same position in an image, which is rotated $180^\circ$ around the centre of the galaxy. To compute Asymmetry, we need a known position of the centre of the galaxy as well as  well-defined area. 
\cite{Abraham96a,CAS} apply a further correction to remove a contribution to the Asymmetry value by the sky background. See section \ref{sss:sky} why we chose not to in this paper.

\subsubsection{Smoothness}
\label{sss:S}

Following \cite{Takamiya99}, \cite{CAS} introduced Smoothness, which has gone through several definitions. Here we use:
\begin{equation}
S = {\Sigma_{i,j} | I(i,j) - I_{S}(i,j) | \over \Sigma_{i,j} | I(i,j) | }
\label{eq:s}
\end{equation}
\noindent where $I_{S}(i,j)$ is the same pixel in a smoothed image. Smoothness is a parameterized version of the unsharp masking technique \cite{Malin78b} used on photographic plates to identify faint structures. The various definitions employ different smoothing kernels and sizes, the most recent one using a flexible kernel-size of 0.2 Petrosian radius and the  boxcar shape. To simplify matters, we chose to use a fixed 6" Gaussian smoothing for our definition. \cite{Abraham96a,CAS} apply a further correction for a background contribution to this parameter, which we do not (See section \ref{sss:sky} as to why).

\subsection{Gini and $\rm M_{20}$}
\label{ss:gm20}

The Gini and $M_{20}$ parameters were established by \cite{Lotz04} as an alternative to the CAS space. 

\subsubsection{Gini}
\label{sss:G}

The Gini parameter is an established qualifier in economics for the inequity in income for a population. 
For a Gini value of one, every person (or pixel) owns an equal fraction of the wealth (or flux). A Gini value of zero, all wealth (or flux) is concentrated in a single person (or pixel). \cite{Abraham03} introduced the Gini parameter, and \cite{Lotz04} used this scale-invariant parameter to characterize the homogeneity of a galaxy image. This parameter shares some of the characteristics of Concentration and Smoothness from the CAS space but does not depend on the size and shape of a convolution kernel or the choice of the galaxy's centre. 
In \cite{Lotz04}, their equation 6, following the work by \cite{Glasser62}, the Gini parameter can be redefined for speed if one orders the pixels according to value first:
\begin{equation}
G = {1\over \bar{I} n (n-1)} \Sigma_i (2i - n - 1) | I_i |\\
\label{eq:g}
\end{equation}
where $I_i$ is the value of pixel i in the ordered list, $n$ is the number of pixels in the galaxy image, and $\bar{I}$ is the mean pixel value in the image. We implemented this definition of the Gini parameter as the computationally least costly one.

\subsubsection{$\rm M_{20}$}
\label{sss:m20}

\cite{Lotz04} also introduce the relative second order moment of the image to classify galaxies. 
The second order moment of a pixel is: $M_i = I_i [(x_i - x_c)^2 + (y_i - y_c)^2]$, where $I_i$ is the value of pixel i in the image, and $x_i$ and $y_i$ are the x and y coordinates of that pixel and $x_c$ and $y_c$ are the position of the galaxy's centre. The total second order moment of an image is defined as:
\begin{equation}
M_{tot} = \Sigma M_i = \Sigma I_i [(x_i - x_c)^2 + (y_i - y_c)^2].
\label{eq:mtot}
\end{equation}
When we now rank the pixels by value, we can define the relative second order moment of the brightest 20\% of the flux:
\begin{equation}
M_{20} = \log \left( {\Sigma_i^k M_i  \over  M_{tot}}\right), ~ {\rm for ~ which} ~ \Sigma_i^k I_i < 0.2 ~ I_{tot} {\rm ~ is ~ true}.\\
\label{eq:m20}
\end{equation}
\noindent Pixel $k$ is the pixel marking the 20\% point in the list of  pixels ranked by flux value.

Some authors vary the central position ($x_c,y_c$) to minimize this parameter \citep{Lotz04,Bendo07}. 
Because we have dynamical centres, we fix $x_c,y_c$ and treat deviations from this value as a source of uncertainty.

\subsubsection{\label{ss:E}Ellipticity}
\label{sss:E}

\cite{Scarlata07} added the ellipticity of a galaxy's image to the mix of parameters in order to classify galaxies according to type in the COSMOS field.  Ellipticity is defined as:
\begin{equation}
E = 1 - b/a
\label{eq:e}
\end{equation}
with $a$ and $b$, the major and minor axes of the galaxy, computed from the spatial second order moments of the light along the x and y axes of the image in the same manner as Source Extractor \citep{se,seman}.

\subsection{Uncertainty Estimates}
\label{ss:unc}

The sources of uncertainty in the above parameters are: (1) shot noise in the pixel values, (2) uncertainties in the position of the centre of the galaxy and (3) variations in the area over which the parameters are computed. 
The first two uncertainties can be estimated using a Monte-Carlo set of iterations, the last one using a jackknifing technique. The relative contribution of these sources of uncertainty depends on the instrument characteristics and hence wavelength, resolution and distance of the object. For instance, photon shot noise is more of an effect in UV and optical data, compared to the \hi. Here however, we treat each image the same and compute these uncertainties for the relevant parameters. 

The shot noise effect on a parameter can be estimated by reassigning random pixel values around the mean value to each pixel in the image and recomputing the parameter several times. With a few iterations, the rms of the spread in parameter values is an estimate of uncertainty in the parameters. 

Similarly, the uncertainty due to the measurement error of the centre of the galaxy ($x_c,y_c$) is can be estimated by varying the input centre, recomputing the parameter and calculating the spread in values over a certain number of iterations (ten in our case). We use random deviations from $x_c$ and $y_c$ within a normal distribution with a width of 6" to mimic the uncertainty in the position of the centre. 

The latter uncertainty estimate is important for Concentration, Asymmetry, $M_{20}$, and Ellipticity as these depend on the assumed centre of the galaxy. To minimize our differences with dynamical parameters later, we adopted the dynamical centre from \cite{Trachternach08} whenever possible. In previous optical high-
ft studies, the uncertainty in the position of the centre was considered less of an issue --since it is generally much less than a resolution element-- but we find that for nearby galaxies, this is a substantial part of the error-budget in the parameters. 

The Gini parameter does not depend on the central position and it is only weakly sensitive to shot noise in the pixel values. 
Therefore, we computed its uncertainty from a shot noise Monte-Carlo set and the rms in Gini values from a series of subsets of the pixels in the image (a jackknife set). The jacknife approach to the Gini parameter uncertainty is advocated by \cite{Yitzhaki91}. Gini shot noise and jacknife uncertainty estimates are of similar magnitude.

The reported uncertainties in Tables (A3-A30, {\em online edition})
are the combined uncertainty from shot noise and central position for all parameters, with the exception for the Gini parameter. In the case of Gini, the uncertainty is the combination of shot noise estimate and the jacknife estimate. These are formal errors and in our opinion, the actual errors in these parameters are larger, predominantly due to viewing angle, resolution effects, image artifacts, and remaining uncertainty about which pixels to include in the computation.

\subsection{Systematics}
\label{ss:sys}

There are several effects that may influence the structural parameters: (1) the sky background, notably any structure in this background, (2) signal-to-noise, (3) the inclination of the disk and (4) the resolution and sampling of the instrument or conversely the distance of the galaxy. These systematics vary in prominence for each different wavelength but we focus on the effects on \hi \ morphology here.

\subsubsection{Background Contribution}
\label{sss:sky}

The sky and instrumental background in these images influences the morphological parameters in three 
different instances. First, the choice of area over which the parameters are computed is influences if a s/n criterion is handled for the inclusion of pixels. Secondly, the {\em mean} sky background over the area contributes to the total parameter value. This is a de-facto weighting of the shape contribution to the parameter. And thirdly, in the case of Asymmetry and Smoothness, the noise in the background adds value since these parameters use the absolute difference in pixel values. 

The first consideration does not come into play in our experiment as we fixed the choice of area based on the \hi \ contour. 
The contribution by the mean background is minimized in our case as we selected uniform data, which is already sky-subtracted, 
and we carefully sky-subtracted the SDSS tiles before addition. Especially the SINGS data was meticulously 
corrected for background contributions \citep[see the fifth SINGS data release notes,][]{dr5}. 
The mean background contribution is dominated by the shape over which the parameters are computed. To compare, we include the parameters computed for the shape with all pixels set to unity in Appendix A ({\em online version of the paper}, Figure A1--A27). 
This is effectively a gross overestimate of a {\em mean} background contribution. This leaves the noise term in Asymmetry and Smoothness as a separate problem.

\cite{Lotz04} correct their Asymmetry and Smoothness values for a background contribution. 
Because these parameters use a absolute difference between pixels, any background noise --or 
other background contribution-- will add to the signal in both these parameters. To correct for 
this, most authors estimate the background contribution in an area of similar size as the object near or 
around the object. In the case of \hi \ maps, this would be complicated by the fact that the \hi \ map covers 
most of the field of view in the other wavelengths.

The reason we did not choose to correct our parameters for a background contribution this way was because the annulus where the 
background contribution would be computed for the inner contour would fall within the outer \hi \ 
contour. Hence, the background may contain signal we are trying to measure. Using an annulus 
further out of the target galaxy is complicated by a lack of coverage in the optical and Spitzer images. 

\begin{figure*}
\centering
\includegraphics[width=0.49\textwidth]{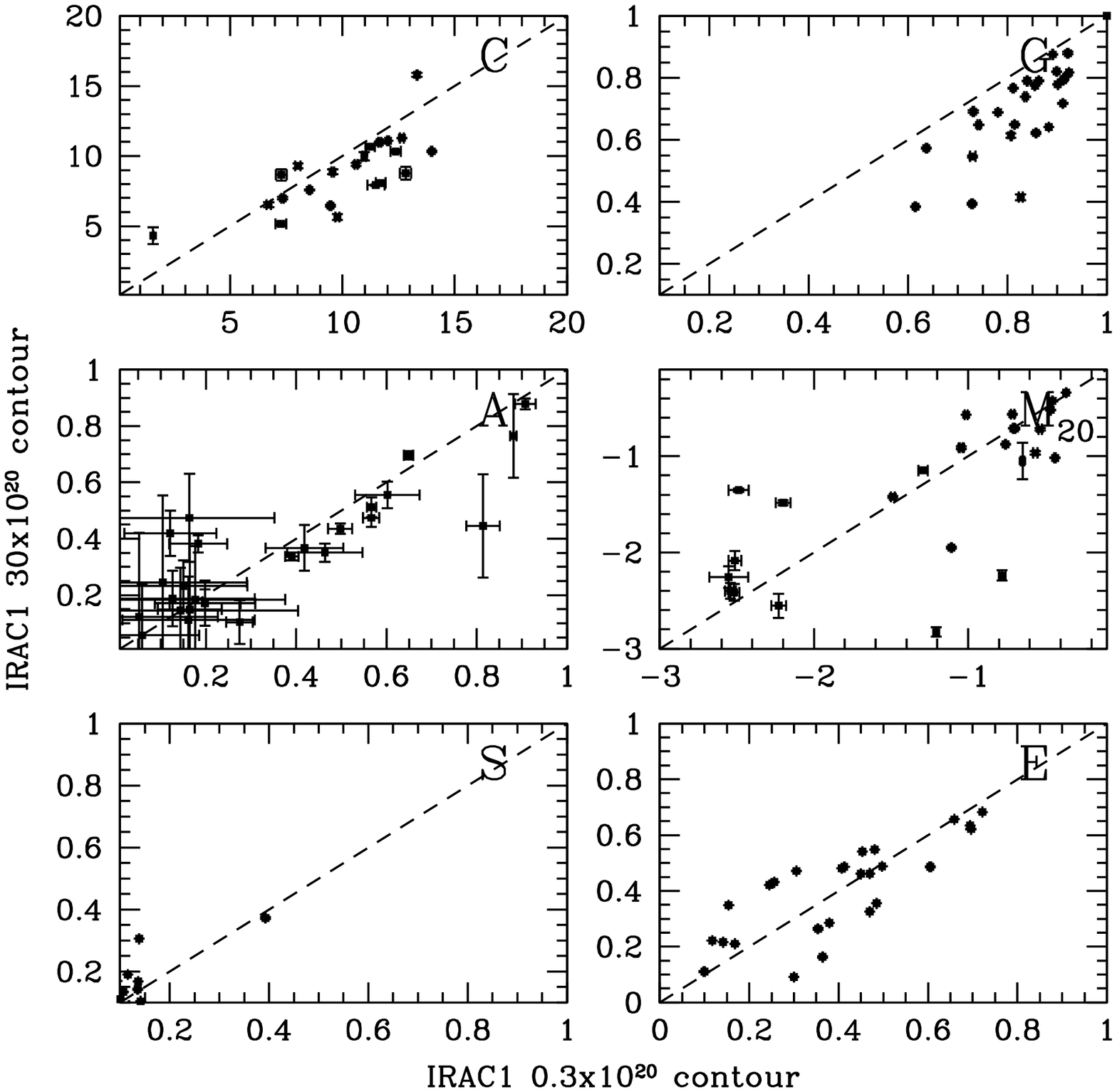}
\includegraphics[width=0.49\textwidth]{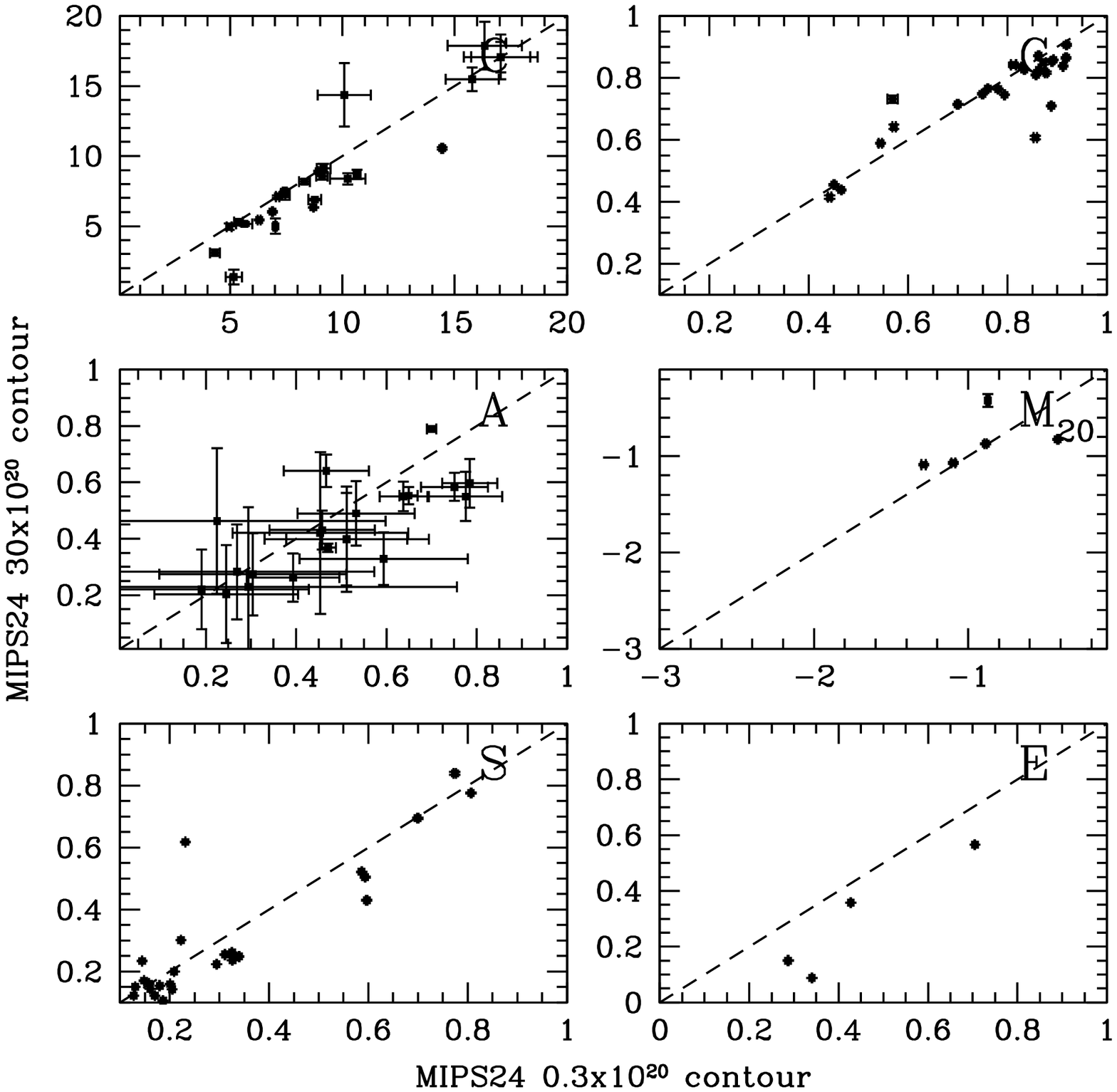}
\caption{The Morphological parameters for the smoothed IRAC1 (3.6 \mum) and native resolution MIPS (24 \mum)  images for the gas disk contour and stellar disk contour. If there is a significant contribution by  a background noise to any of the parameters, this should manifest itself as an offset between the two contours.    }
\label{f:2cont}
\end{figure*}

To test our assertion that the background contribution, which includes any flux by a background 
source that was not masked, is minimized, we compare the parameters in both contours in IRAC1 (3.6 \mum) 
and MIPS 24 \mum \ images, two filters for which we have measurements for all our galaxies. 
The 3.6 images were smoothed to 6" resolution, correlating the noise which would strengthen 
any weak background patterns and the MIPS 24 was not smoothed but a likely candidate for 
background structure. 
Figure \ref{f:2cont} shows the comparison between the two contours. While there is much scatter, there is 
no systematic offset in either Asymmetry or Smoothness. Similar plots for the other wavelength show a similar trend. 
This represents a lower limit to a background structure contribution from for example instrument characteristics etc. 
However, the lack of an offset is encouraging.
The shot noise uncertainty (\S \ref{ss:unc}) is separate effect from any background contribution but we note that, 
especially in Asymmetry, any background deviations seems to fall within the computed uncertainty.

Alternatively, one could construct background images, mimicking the background mean value and noise. 
However, if these do not include a mean background contribution, there is very little change in the values 
for all morphological parameters (at least for the 3.6 and 24 \mum images where we tried this). 
The large signal-to-noise over many pixels in the selected shape dominates the parameter value.

Therefore, we do not correct our values for a background contribution here. However, these parameters have 
been determined for a large number of pixels and high signal-to-noise images of nearby galaxies. 
As soon as the background can be a substantial contribution to the intensities in an image, such is 
the case in HST images of distant galaxies, the background contribution needs to be revisited.
In the case of \hi \ observations in particular, this issue will be much less prominent as the column density 
map is constructed from high signal-to-noise line detections, lowering the relative contribution of the background.

\subsubsection{Signal-to-Noise}
\label{sss:sn}

\cite{Lotz04} also conduct an experiment how much these parameters are affected by a change in signal-to-noise. They find that G, $M_{20}$, and C are reliable to within 10 \% for galaxy images with S/N $>$ 2 per pixel. \cite{Lisker08} find a similar signal-to-noise limit for the Gini parameter. Asymmetry and Smoothness decrease with S/N but stay within limits above S/N $>$ 5. The stellar disk of all our galaxies are detected well beyond this noise level. The \hi \ perspective will initially be used in the nearby universe where the signal-to-noise level will comfortably exceed these limits because line detections exceed s/n of 3 or 5 in individual channel maps to be included in the column density map. Signal-to-noise effects will need to be considered when the classification scheme is applied to galaxies at higher redshifts with for instance SKA. 
The redistribution of pixel values around the mean (the Monte-Carlo error estimate) in our parameter is in part to quantify how much effect noise has on our parameter determinations.

\subsubsection{Distance Effects on Morphology}
\label{sss:rez}

The accuracy of the morphological parameters depends on sampling and resolution of the analysed images. In addition, cosmological surface brightness dimming plays a role at greater distances. 
Here, we focus on the effects of distance on the morphological parameters measured in our \hi \ column density maps.
In this section, we focus on the effects on increasing distance on the morphological measurements in our \hi \ column density maps, to ensure that these THINGS measures do not need to be corrected for relative distance.

Distance effects in the optical and infrared images have been discussed by both \cite{Lotz04} and \cite{Bendo07}.  \cite{Lotz04} find that decreased sampling has the strongest effects on Concentration and $M_{20}$; 15\% changes when pixel scales become greater than 500 pc. Gini, Asymmetry, and Smoothness, on the other hand, are relatively stable with decreasing spatial scales down to 1000 pc. In the near and far-infrared, \cite{Bendo07} find that Concentration, Gini and near-infrared Asymmetry ($A_{3.6\mu m.}$) are invariant with image smoothing and the $M_{20}$ parameter is only moderately affected ($<$20\%). The one exception is the Asymmetry in the far-infrared (24 \mum). They report a dramatic change with distance; $A_{24 \mu m.} \propto d^{-0.26}$, with $d$ in Mpc. 
The spatial resolution of these 24 \mum \ images is lower than the images at other wavelengths. However, they do sample smaller physical scales than the limit 1 kpc found by \cite{Lotz04} for the applicability of these parameters. The dramatic change in Asymmetry found by Bendo et al. seems to be somewhat in contradiction to the smooth decrease found by Lotz et al. 
The general picture is that as long as features smaller than one kiloparsec are resolved in the images, the morphological parameters do not suffer too much but decline rapidly for coarser images.

The THINGS observations are designed to sample the \hi \ at spatial scales below a kiloparsec but the future all-sky surveys will most likely sample the \hi \ disk at lower spatial resolutions. To quantify the effects of distance on the morphological measurements, we use the \hi \ RO maps of five face-on spiral galaxies: NGC 628, NGC 3184, NGC 3351, NGC 5457 and NGC 6946. To shift these to a series of distances, we cosmologically dimmed the maps, smoothed with 6" resolution at the appropriate distance, and resampled to 1" pixel scale of the THINGS survey, adding in the noise reported by \cite{Walter08} for a single channel map (their table 2)\footnote{Our \hi \ maps are observed throughout the datacube but we treat these maps as single images, similar to the redshifting prescription from \cite{Giavalisco96}.}. 

\begin{figure}
\centering
\includegraphics[width=0.49\textwidth]{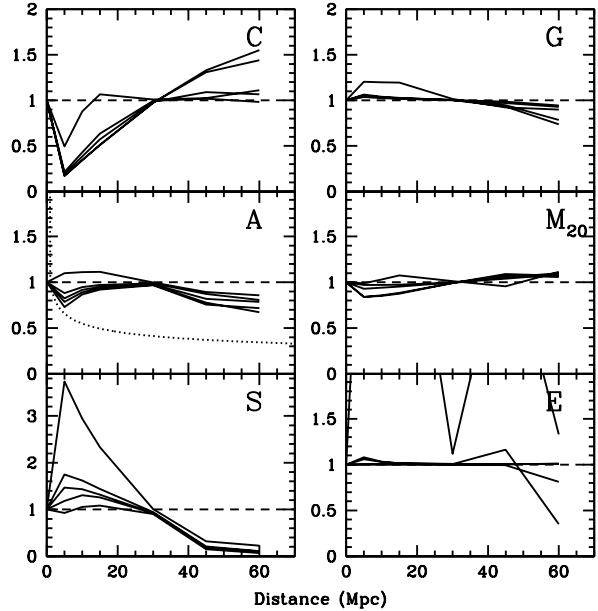}
\caption{The effect of distance on \hi \ galaxies. All morphological parameter are expresses as the ratio between the original value and the one found at greater distance. All five galaxies were dimmed, smoothed with the resolution of THINGS (6") and regridded to the THINGS pixel scale (1").  The correction for Asymmetry with distance from \protect\cite{Bendo07} is marked with a dotted line (for Spitzer/MIPS data with 6" resolution).   }
\label{f:dist}
\end{figure}

In Figure \ref{f:dist} we plot the ratio of the measured morphological parameter over the original, fiducial values (RO maps at the fiducial distance of zero).

Concentration and Smoothness of the \hi \ maps are affected the most by distance effects. In the case of Smoothness, this is unsurprising as it relies on a smoothed version of the original image for the measurement. In the optical, the size of the smooting kernel is adapted to be sensitive to specific scales of star-formation. Hence, additional smoothing with distance is very likely to affect this value. 

Asymmetry, Gini, and \m20 change less than 20\% in value with increasing distance. This is encouraging for their use in throughout a local volume survey. 
Asymmetry certainly does not show the kind of behaviour described by \cite{Bendo07} for the 24 \mum \ SINGS maps (dotted line in Figure \ref{f:dist}). Considering the 24 \mum ~ images trace the dusty star-formation and our maps the cool atomic gas, this is not as surprising as it initially appears. In the 24  \mum \  maps, a large fraction of the flux is in a few star-forming regions, dominating morphological measurements. In the case of \hi, the morphology is determined by a large area with a very limited range of column densities. The difference in contrast for these observations may account for their different behaviour with increased distance.
 Ellipticity does not change in value much either. These galaxies have very low values of ellipticity as they are all near face-on. 

In a companion paper in this series \citep{Holwerda10d}, we quantify the effects of increased distance using simulated \hi \ maps (initially at 150 pc. sampling) and typical resolutions, pixels-scales and distances for current and planned surveys.
From these simulated \hi \ maps, it becomes clear that with current and planned observatories, morphological \hi \ measurements are only useful for the very local volume. 


\subsection{Inclination Effects}
\label{ss:incl}

\begin{figure}
\centering
\includegraphics[width=0.5\textwidth]{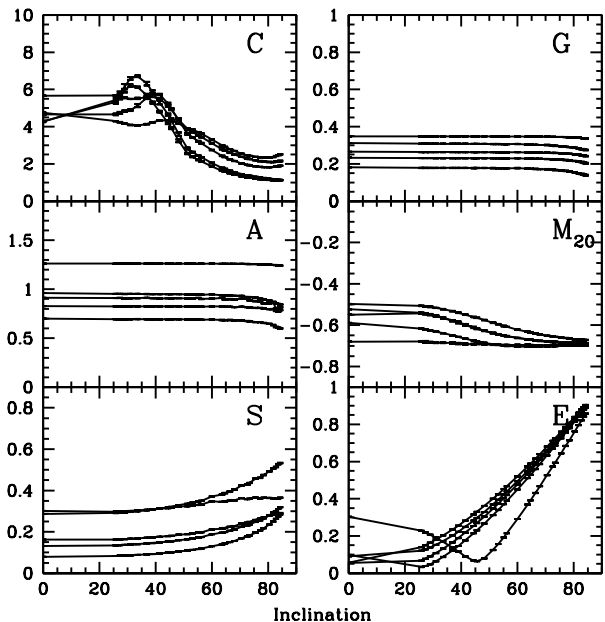}
\caption{The changes in the values of the morphological parameters as a function of inclination of the disk. NGC 628, NGC 3184, NGC 3521, NGC 5457 and NGC 6946 were used. We assume that the \hi \ disks are at zero inclination at the beginning and we rotate around the Y-axes.  }
\label{f:incl}
\end{figure}

In the case of the Concentration defined by two circular apertures, the inclination dependence is an accepted feature, However, even in the case of elliptical apertures, there is often a remaining concentration-inclination relation. For example \cite{Bendo07} find that inclination influences the measured Concentration and they introduce a correction. Other parameters may well have subtle dependencies on the inclination of the disk. 
To explore these dependencies, we use the same five face-on galaxies as for the resolution simulations and rotate them around the x-axis assuming an infinitely thin disk for the \hi \ emission (resampling solely along the y-axis.). While these galaxies have some intrinsic inclination (see Table A1 in the Appendix {\em electronic version of the manuscript}), 
we treat them as perfectly face-on as a starting point.
Figure \ref{f:incl} shows all six parameters as a function of inclination. A benefit of \hi \ observations is a good new estimate of the inclination from the tilted ring fit to the velocity cube. Any correction for inclination will consequently be easy to perform for low redshift galaxies. 

Asymmetry, Smoothness and Gini are all relatively unaffected below an inclination of 80$^\circ$, above which our assumption of a perfectly flat disk becomes tenuous. Ellipticity naturally changes drastically by inclining the disk. $M_{20}$ converges to a value of -0.7. 
Concentration follows a complicated pattern peaking in the $30--40^\circ$ range. \cite{Bendo07} find a much smoother correction of concentration for inclination, based on their elliptical apertures. Arguably, they pre-corrected for inclination by using an elliptical apertures whose shape depends on the inclination of the disk. 

\section{Results}
\label{s:results}

Appendix A ({\em online version of the paper}) shows the column density maps, plots of the parameters over wavelength and lists morphology parameter values for each of our \nsamp \ galaxies. The column density map of each galaxy shows the two areas in the images, corresponding to the extent of the gas (blue) and stellar (red) disk, defined by us as the 0.3 and $30 ~ \times ~ 10^{20}$ cm$^{-2}$ \hi \ contours respectively (or alternatively 0.24 and 24 $M_\odot/pc^2$ respectively). We exclude those parts of the images where the \hi \ map is below these values, even in the case of \hi \  ``holes'' in the \hi disk. 
All six parameters, determined within these contours are plotted for each wavelength we obtained data for. In order to compare different wavelengths, we defined the areas over which all the morphological parameters were computed exclusively based on the \hi \ column density map. Appendix A ({\em online version of the paper}) also lists our brief notes on what we found for each galaxy in the literature, especially regarding its level of interaction. First we will compare our parameters to previous morphological results and secondly to other measures of interaction. 

\subsection{Comparison to Previous Results}
\label{ss:comp}

The SINGS sample, and hence the THINGS sample, has some overlap with the sample from \cite{Frei96}. \cite{CAS} presents CAS parameters for the Frei sample, and \cite{Lotz04} presents Gini, $M_{20}$ and CAS parameters for the Frei sample, both in the optical R-band filter. \cite{Bendo07} presents Concentration, Asymmetry, Gini and $M_{20}$ for two infrared wavelengths for the entire SINGS sample. Here, we compare our results for the \hi \ contour corresponding to the stellar disk.

\subsubsection{Optical}
\label{sss:opt}

\begin{figure}
\centering
\includegraphics[width=0.5\textwidth]{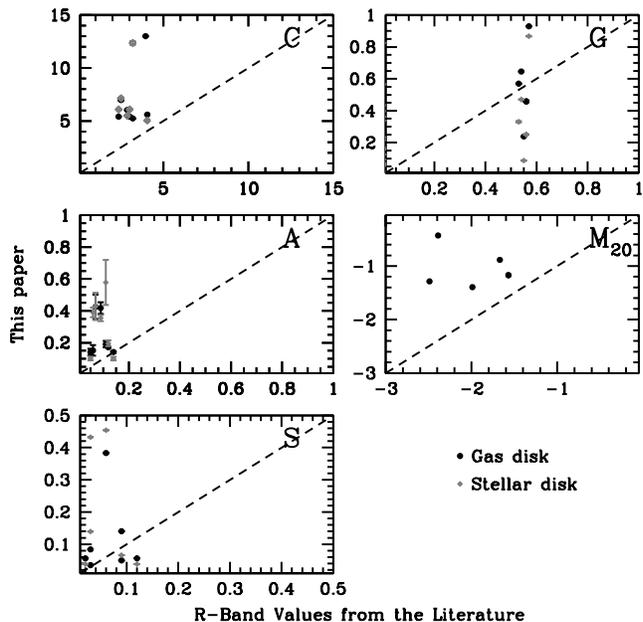}
\caption{\label{f:frei} The R-band values for CAS from \protect\cite{CAS} (left panels) and \protect\cite{Lotz04} for the Gini and $M_{20}$ parameters (right panels).  }
\end{figure}

\cite{CAS} present Concentration, Asymmetry and Smoothness in the R-band for the Frei sample, and \cite{Lotz04} presents separate estimates of the Concentration-Asymmetry-Smoothness values as well as Gini and $M_{20}$ for the Frei sample, in both R and B band. The measurements from these two papers are the nearby (isolated) galaxy reference for CAS measurements. In Figure \ref{f:frei} we compare the \cite{CAS} and \cite{Lotz04} values respectively to ours for the R-band values for our ``stellar" disk (the $30 \times 10^{20}$ cm$^{-2}$ contour). 
The main difference between our implementation and the earlier work is the definition of the area over which the parameters are computed; we define the area over which the parameters are computed based on an \hi \ contour. In contrast, both previous authors use an isophote in the optical or near-infrared image\footnote{Optical isophote and surface density contour are technically identical terms. We chose to use contour to emphasise that the choice of area is done based on the \hi \  column density map, not the optical images.}. The Frei et al. images and the SINGS R-band ancillary data are not exactly the same in depth and filter which may result in additional differences.

There are some small differences in the parameters's definitions as well. Both \cite{CAS} and \cite{Lotz04} minimise the Asymmetry by allowing the galaxy's centre to vary. \cite{Bendo07} and \cite{Lotz04} have a similar approach to $M_{20}$. We chose to use the dynamical centre from \cite{Walter08} and treated variations on it as an uncertainty. 

The Gini values from \cite{Lotz04} span a much smaller range that our values or those from \cite{Bendo07}, possibly because of the difference in area definition. 
Given that our areas are greater than those in Lotz et al., there is a relatively bigger contribution by a mass of low-intensity pixels, increasing the Gini parameter. Similarly, Bendo et al. compute the Gini value for a greater area (the RC3 ellipse), including many more low-flux pixels. This difference in area would explain higher values of Gini compared to Lotz et al., but not necessarily why there is a bigger spread in our values.  
In addition, Bendo et al. note that AGN activity change the IR values considerably and exclude the centre for some of their galaxies for this very reason.
A central AGN is likely to be excluded automatically by an \hi \ contour. 
Our concentration values for R-band may well be higher is some central pixels are excluded by the \hi \ contour (e.g., central \hi \ hole in NGC 2841, Figure \ref{f:n2841}).
The difference in the $M_{20}$ values is mostly due to the differences in area or possibly the depth of the data.

\subsubsection{Infra-red}
\label{sss:spitzer}

\cite{Bendo07} present values for Concentration, Asymmetry, Gini and \m20 for the 3.6 and 24 \mum \ images in the SINGS sample and by design there our sample is a subset of this sample. \cite{Munoz-Mateos09a} compare these infrared values to those in the UV. Both authors focus on their use for Hubble Type classifiers. \cite{Bendo07} correct their Asymmetry values for distance and Concentration values for inclination. The other significant difference between our determination and that of Bendo et al. is that their result is from an elliptical aperture \citep[defined by the de Vaucouleur's $D_{25}$ in the RC3][]{RC3}, and ours are for an \hi \ contour.

\begin{figure*}
\centering
\includegraphics[width=0.49\textwidth]{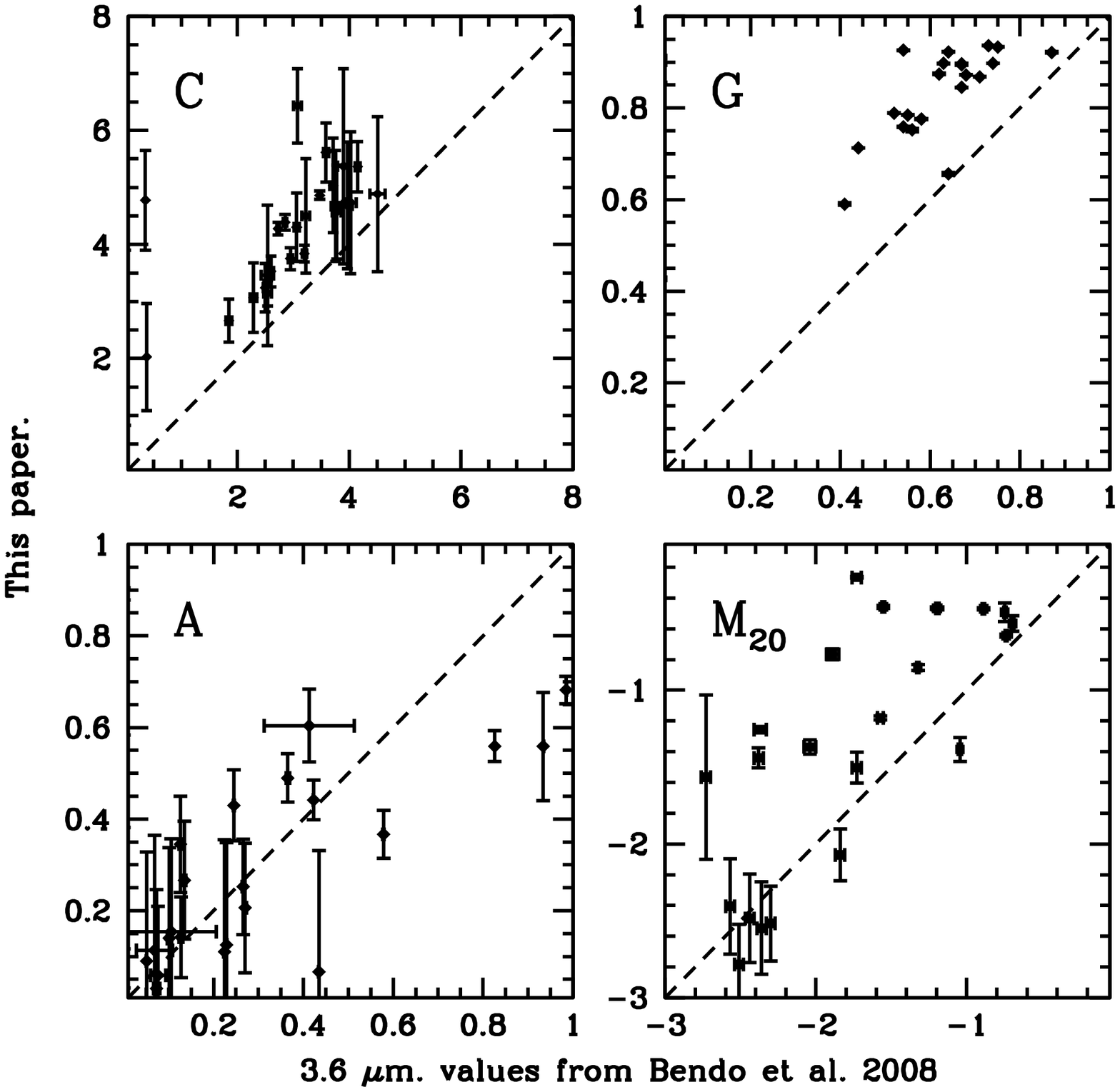}
\includegraphics[width=0.49\textwidth]{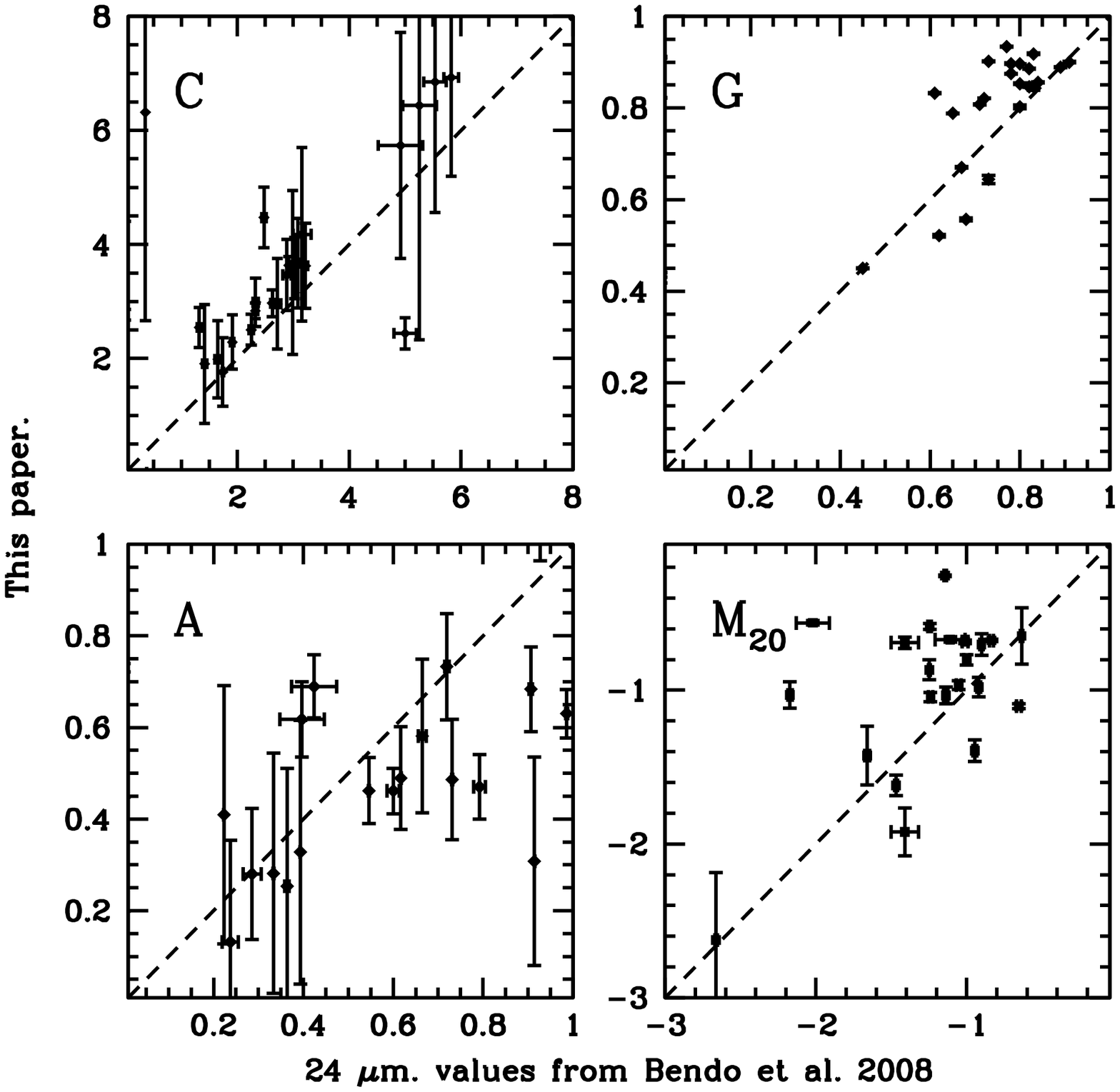}
\caption{Our measurements of Concentration, Asymmetry, Gini and $M_{20}$ in an elliptical aperture defined by the major and minor axes ($D_{25}$) from NED and the RC3 position angle for both the 3.6 and 24 \mum \ images, compared to the values from \protect\cite{Bendo07}. The dashed line is the line of equality.}
\label{f:bendo}
\end{figure*}

Therefore, we recomputed our implementation of the morphological parameters (Concentration, Asymmetry and Gini  and $M_{20}$) for the elliptical apertures used in \cite{Bendo07}. We placed the apertures centered on the galaxy position from \cite{Walter08}, using the minor and major axes from the NASA Extragalactic Database (\url{http://nedwww.ipac.caltech.edu/}) and the Reference Catalog \citep[RC3][]{RC3} position angle. Figure \ref{f:bendo} shows the values from \cite{Bendo07} and ours. There is scatter in the values but that is to be expected as the input apertures, central positions and data are not perfectly identical. 
Concentration and Asymmetry are corrected by Bendo et al. for disk inclination and distance respectively. We have uncorrected these values here for the comparison. 

Our 3.6 \mum \  Concentration and Gini values are higher than those from Bendo et al. In past, this may be because Bendo et al. remove the contribution by the AGN in a few of their galaxies, as well as small differences in the adopted centre of the disk. There is substantial scatter in the $M_{20}$ to higher values for the 3.6 \mum \  and our Asymmetry values, in both 3.6 and 24 \mum \ are lower for the highest values for Asymmetry from Bendo et al. These differences are because we smooth the 3.6 \mum \ data to 6" resolution. 
In addition, some of the differences in $M_{20}$ and Asymmetry can be the result of slight differences on which pixels on the edge of the aperture are included in the computation. Both of these parameters weigh pixels in favor of those on the edge of the object 

All considering, the values for these four parameters agree with the previous authors within our computed uncertainties. Smoothing effect should be kept in mind when comparing this paper's values to others but the smoothing is essential for the comparison across wavelength.

\begin{figure}
\centering
\includegraphics[width=0.5\textwidth]{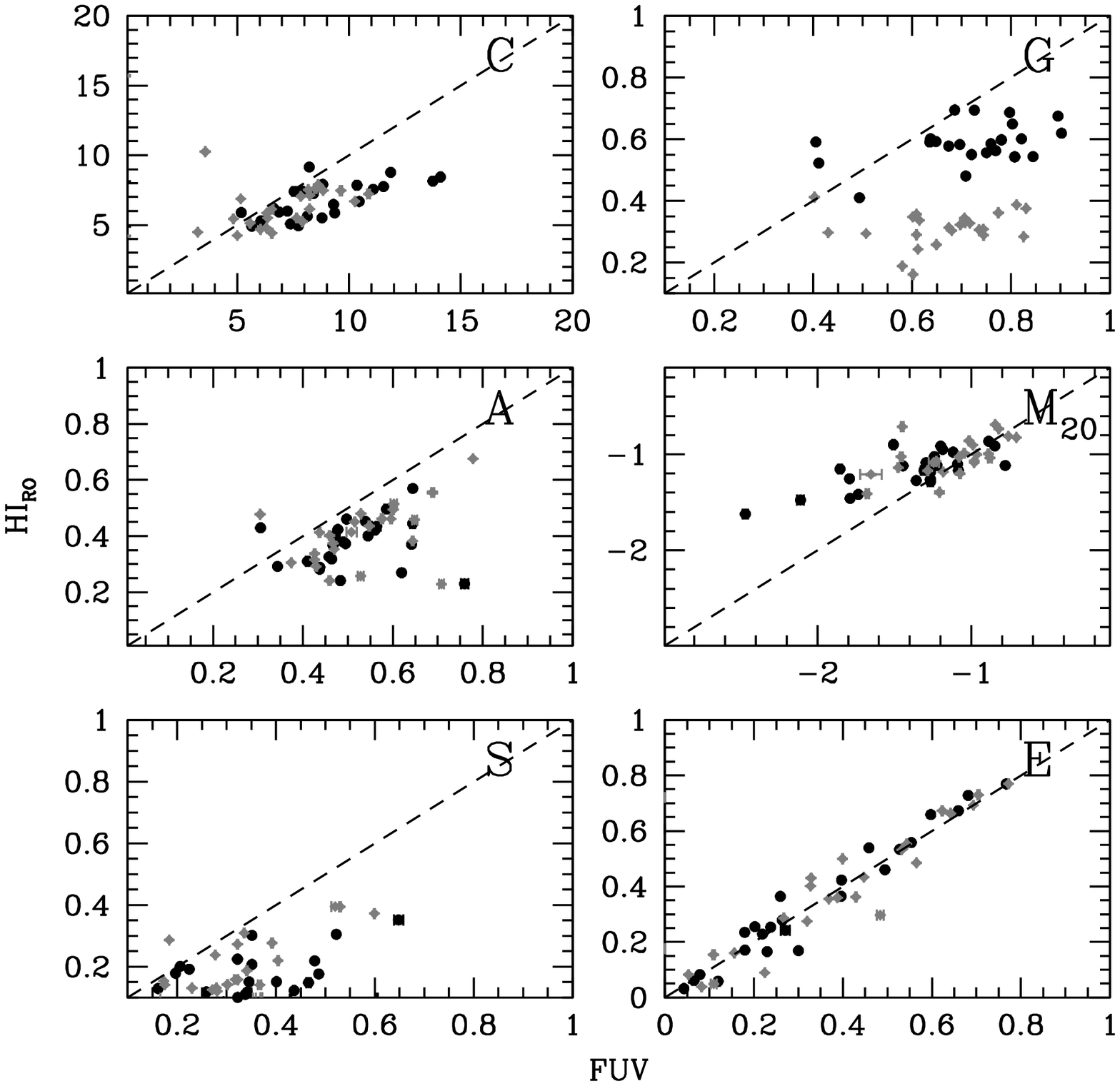}
\caption{Our measurements of Concentration, Asymmetry, Smoothness, Gini, $M_{20}$ and Ellipticity in both the stellar disk ($3 \times 10^{21}$ cm$^{-2}$, solid points) and the gas disk contour ($3 \times 10^{19}$ cm$^{-2}$, grey points) in far-ultraviolet compared to the values computed in the \hi \ RO map.}
\label{f:fuv}
\end{figure}

\begin{figure}
\centering
\includegraphics[width=0.5\textwidth]{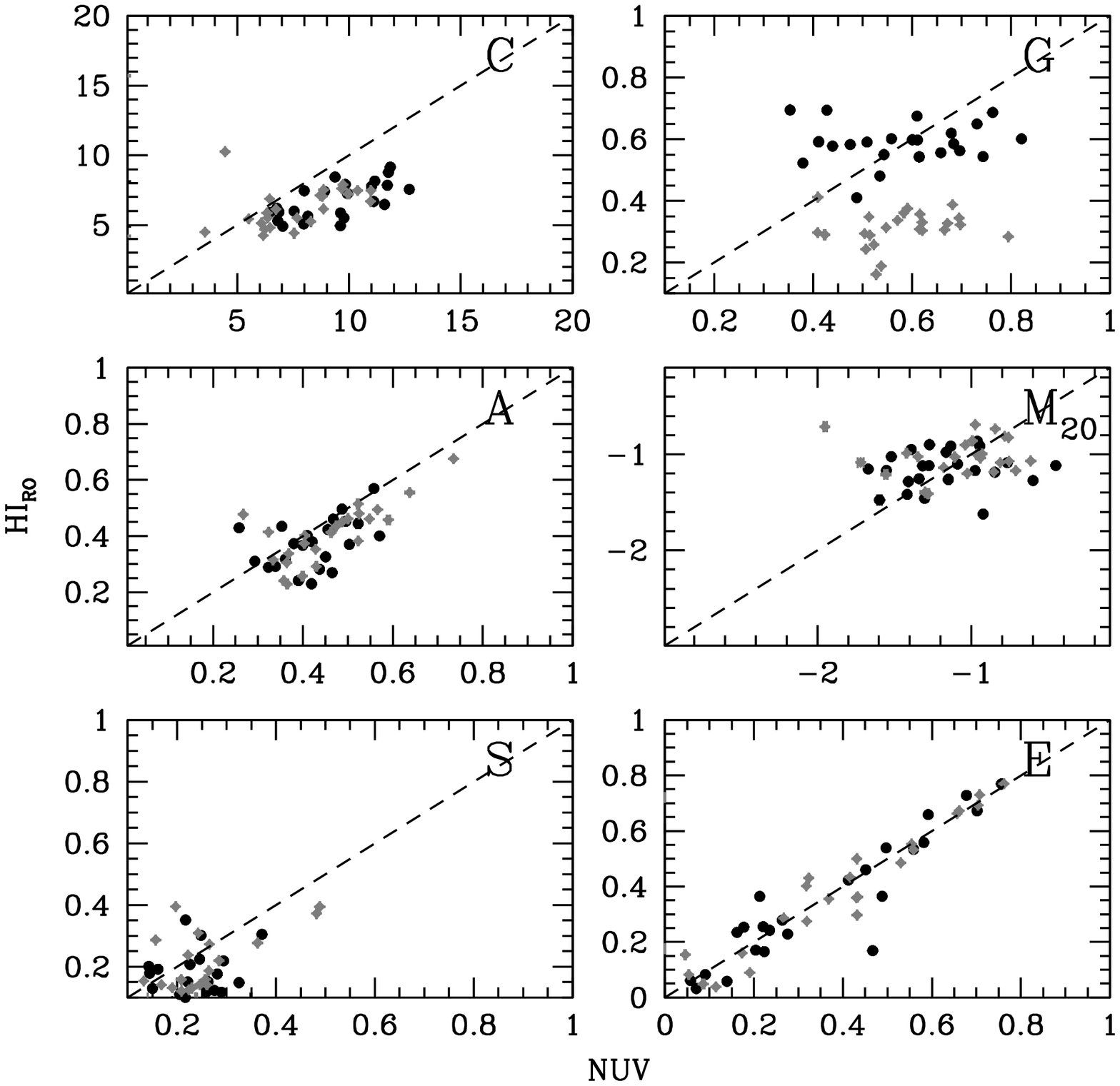}
\caption{Our measurements of Concentration, Asymmetry, Smoothness, Gini, $M_{20}$ and Ellipticity in both the stellar disk ($3 \times 10^{21}$ cm$^{-2}$,solid points) and the gas disk contour ($3 \times 10^{19}$ cm$^{-2}$, gray points) in near-ultraviolet compared to the values computed in the \hi \ RO map.}
\label{f:nuv}
\end{figure}

\begin{figure}
\centering
\includegraphics[width=0.5\textwidth]{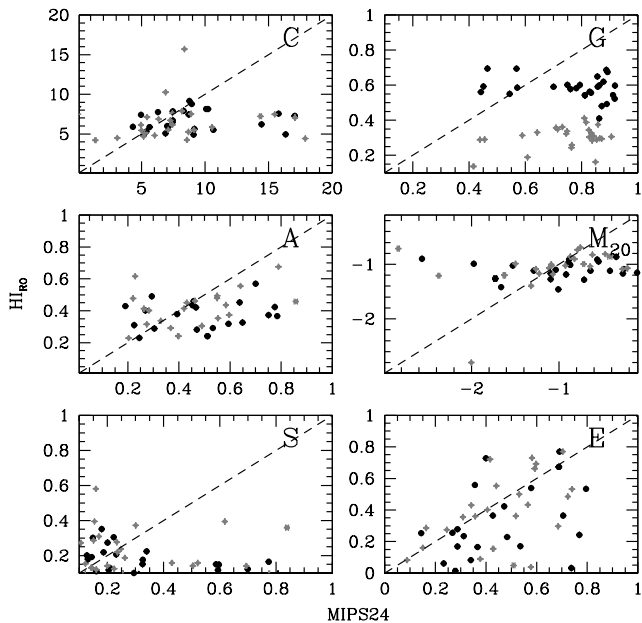}
\caption{Our measurements of Concentration, Asymmetry, Smoothness, Gini, $M_{20}$ and Ellipticity in both the stellar disk ($3 \times 10^{21}$ cm$^{-2}$, solid points) and the gas disk contour ($3 \times 10^{19}$ cm$^{-2}$, grey points) in Spitzer 24 \mum \ compared to the values computed in the \hi \ RO map.}
\label{f:24}
\end{figure}

\subsection{Morphology of Star-Formation and \hi}
\label{ss:comp}

Because most of the existing work has been done in wavelengths dominated by recent star formation, mostly UV, we compare the morphological parameters determined in the \hi \ RO maps to the parameters we determined in the NUV, FUV and MIPS 24 \mum. The spatial resolutions in these wavelengths are all very similar and we use the same two areas; the $3 \times 10^{21}$ and $3 \times 10^{19}$ cm$^{-2}$ \hi \ contours. Figures \ref{f:fuv}, \ref{f:nuv} and \ref{f:24} show the six parameters computed in the far-, near-ultraviolet and 24 \mum \ compared to the values in the RO \hi \ maps. 

Ellipticity translates extremely well from the UV to the \hi. Concentration, $M_{20}$, and Asymmetry also translate well but to a lesser extent. 
However, this can be expected for the \hi \  disk. For instance, it was already known that gas disks are less concentrated, often with depressions in the centre of the galaxies. In the case of the Gini parameter, one could expect a shift because the range in values in an \hi \ column density map is much smaller than the range in an UV flux map. Small changes in Asymmetry and $M_{20}$ can be expected as a bright star-forming region does not necessarily translate into a high \hi \ column density.
Asymmetry for the \hi \ and UV follows each other much better than the 24 \mum \ emission. A similar relation is seen in $M_{20}$ where \hi \ and UV are reasonably similar but the 24 \mum \ is not. Smoothness shows no clear relation. 

In general, the morphological parameters from the far- or near-ultraviolet translate reasonably well to the \hi, with the exceptions (Concentration and Gini) well understood. The discrepancy is much greater with the 24 \mum \ MIPS observations. Thus, \hi \ and restframe ultraviolet merger measurements from morphology translate directly but any FIR morphology requirers additional 
information. Arguably, FUV and \hi \ trace the atomic ISM component and the 24 \mum \ is closer associated with the molecular ($H_2$) component. In favor of this argument is the poor relation in $M_{20}$; the bright pixels in 24 \mum \ do not coincide with high column densities in \hi.
Fortunately, the majority of morphological measurements of mergers are in rest-frame UV and our measurements of merger fractions and rates should translate relatively directly.

\section{Signatures of Tidal Interaction}
\label{s:sig}

In this section we explore the relation between other methods to determine the level of interaction for each galaxy to our morphological parameters. 
The level of interaction is difficult to quantify although some authors present a parameterisation of the level of tidal force on a galaxy \citep{Karachentsev04,Bournaud05b}. Alternatively, one can take the level of non-circular motion from the THINGS results as a measure of recent interaction. Many of these galaxies have been studied extensively and on several, previous authors have remarked signs of disturbance (see Table \ref{t:rank}). In this section, we compare our morphological parameters in \hi \ to these various parameterisations to identify those parameters that appear to be the most promising interaction tracers.

\begin{figure}
\centering
\includegraphics[width=0.5\textwidth]{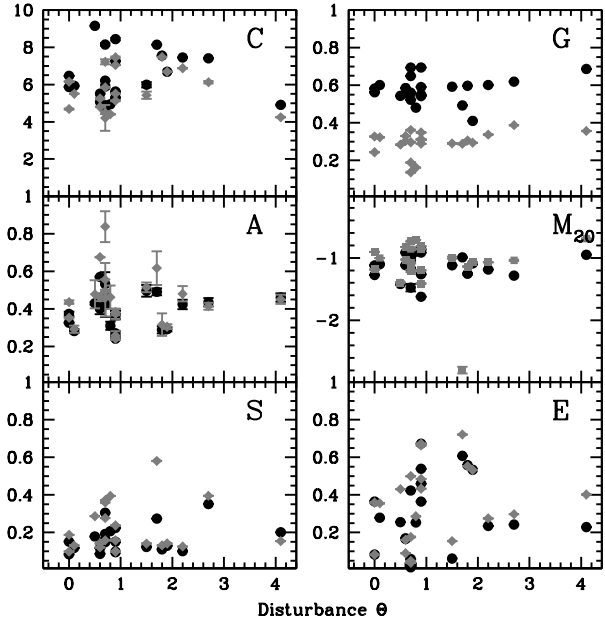}
\caption{\label{f:theta} All six morphological parameters, Concentration, Asymmetry, Smoothness, Gini, $M_{20}$ and Ellipticity as a function of tidal disturbance ($\Theta$) from \protect\cite{Karachentsev04}. Grey points are for the outer contour, (the gas disk) and the solid points for the inner contour (the stellar disk). }
\end{figure}

\subsection{Tidal Interaction Parameter ($\Theta$)}
\label{ss:theta}

One way to measure the interaction rate is to find close galaxy pairs which, most likely, will be gravitationally interacting. 
\cite{Karachentsev04} presents a catalogue of neighbouring galaxies and a tidal estimate for each galaxy ($\Theta$).
Negative values of $\Theta$ correspond to isolated galaxies, and positive values are typical of group members.
We should note that there are no completely isolated galaxies in our sample ($\Theta<0$). Only NGC 628 and NGC 2403 
have $\Theta=0$; galaxies in average surroundings and we note that NGC 2403 was the zero-point calibrator for this tidal parameter. 
Five galaxies have no value for $\Theta$ (NGC 2841, NGC 3198, NGC 3521, NGC 5055, NGC 7331). This makes a 
direct comparison between the morphological parameters and tidal interaction more difficult. Figure \ref{f:theta} shows
no clear relation between any of the morphological parameters of the \hi \ maps and $\Theta$, compared either over the optical and gas disk.

Because we only have one galaxy with a high value of the Karachentsev tidal estimate ($\Theta=4$ for NGC 5194), we 
can only glean trends with this parameter: compared to the locus of group galaxies, the $M_{20}$ parameter for the \hi \ map 
appears to be higher for NGC 5194, and Concentration, Asymmetry and Gini show a slight trend. 
We interpret the relation with $\Theta$ useful to point out those parameters that could be of use to identify mergers in \hi \ 
morphology, but the THINGS sample does not have the spread in $\Theta$ to isolate the \hi \ morphological parameter 
space where gravitational interactions reside.

\begin{table*}
\caption{The ranking of interaction}
\begin{center}
\begin{tabular}{l l l l}
	& Name		& notes							& reference \\
\hline
1. &	NGC 2841	& \hi \ warp						& \\
2. &	NGC 3184 	& 								& \\

3. &	NGC 3521	& 								& \\
4. &	NGC 3621	& 								& \\

5. &	NGC 3198	& strong non-circular motion			& \cite{Trachternach08}\\
6. &	NGC 2903	& small companion 					& \cite{Irwin09}\\

7. &	NGC 628		& high-velocity clouds				& \cite{Kamphuis92} \\
8. &	NGC 925		& \hi \ tail							& \cite{Sancisi08}\\
9. &	NGC 5457	& lopsided						& \cite{Richter94} \\
10. &	NGC 6946	& holes, high-velocity complexes 		& \cite{Boomsma08} \\

11. &	NGC4736		& \hi \ streaming in and lopsided.		& \\

12. &	NGC 7793	& Sculptor Group member, warp		& \\
13. &	NGC 7331	& Proximity to StephanÕs Quintet 		& \cite{Gutierrez02}\\

14. &	NGC 2403	& M81 group member, 				& \cite{Fraternali02}\\
 &				& extra-planar \hi					& \\
15. &	Holmberg II	& M81 group member				& \cite{Stewart00}\\
16. &	M81A		& M81 group member				& \\
17. &	DDO 53		& M81 group member				& \\
18. &	Holmberg I	& M81 group member				& \\
19. &	NGC 2976	& M81 group member				& \\
20. &	IC 2574		& M81 group member, 				& \\
	&			& \hi \ supershell					& \\
21. &	NGC 3031	& M81 central galaxy				& \\

22. &	NGC 3627	& member interacting Leo triplet		& \\
23. &	NGC 5194	& canonical 3:1 interaction			& \\

24. &	NGC 5055	& extended spiral structures, 			& \cite{Bosma81b}\\
				&		& \hi \ warp						& \\
25. &	NGC 4826	& counter-rotating \hi \ disk			& \cite{Braun94} \\
26. &	NGC 5236	& high-velocity clouds, stellar and		& \cite{Malin97b} \\
&				& \hi \ streams and an \hi \ warp.		& \cite{ghosts}\\
\hline

\end{tabular}
\end{center}
\label{t:rank}
\end{table*}%

\begin{figure}
\centering
\includegraphics[width=0.5\textwidth]{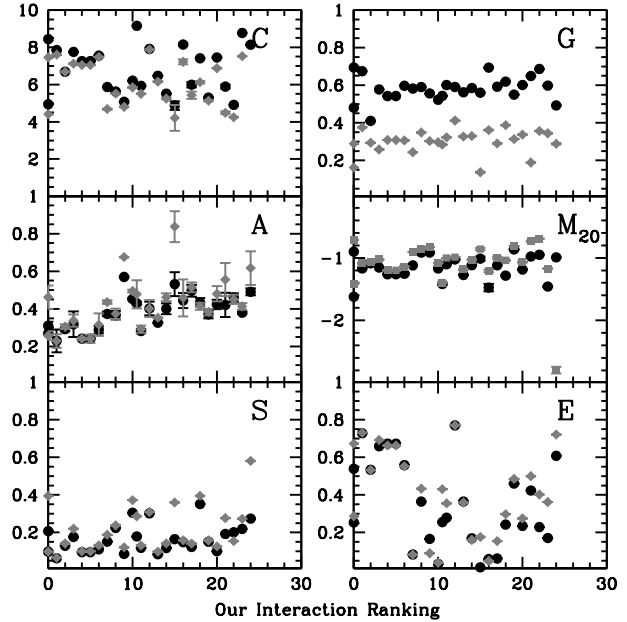}
\caption{\label{f:rank} All six morphological parameters, Concentration, Asymmetry, Smoothness, Gini, $M_{20}$ and Ellipticity as a function of our ranking. Grey points are computed over the gas disk, solid points for the inner stellar contour. We note the marked increase in Asymmetry.}
\end{figure}

\subsection{Interaction Ranking}
\label{ss:rank}

Because the THINGS sample was mostly chosen to include nearby and non-interacting galaxies (M51 being the obvious exception), the majority of the above galaxies are not in the stage of an interaction where the morphological signature is the strongest \citep{Lotz08a,Lotz10a,Lotz10b}. So much so that \cite{Smith07a} use the SINGS sample (minus M51) as their reference for non-interacting galaxies. However, we can rank the THINGS galaxies on the level of interaction signature reported in the literature. In Table \ref{t:rank} we rank the sample based on the plausible stage of interaction from isolated, non-harassed galaxies to merger and merger remnant. 
This ranking is subjective as more information on a galaxy (such as high-resolution \hi \ data) often reveals signs of mild interaction (e.g., warps or low column density structures), and it is still unclear what signatures are from interaction or not. For instance, warps may be formed by other means as well (e.g., IGM ram pressure). Certain features are long lived (e.g., warps) while other fade relatively quickly (tidal arms).
Figure \ref{f:rank} shows all six parameters as a function of the ranking in Table \ref{t:rank}. Asymmetry shows a gradient with the ranking while $M_{20}$ does not. In our opinion, this does not mean Asymmetry is a better parameter to detect interaction, just that it is very sensitive and may well pick up other effects or longer lasting effects as well. 

\begin{figure}
\centering
\includegraphics[width=0.5\textwidth]{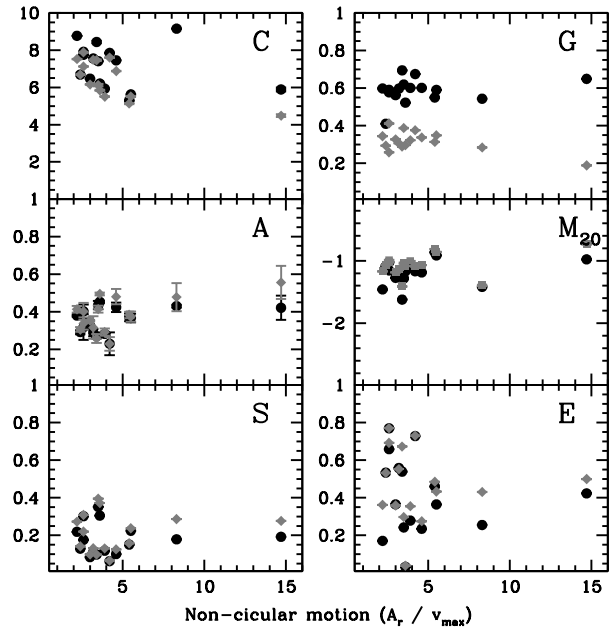}
\caption{\label{f:Ar} All six morphological parameters, Concentration, Asymmetry, Smoothness, Gini, $M_{20}$ and Ellipticity as a function of the non-circularity of the \hi \ kinematics in the disk according to \protect\cite{Trachternach08}. Grey points are the gas disk, solid points are computed for the inner stellar contour.  }
\end{figure}

\subsection{Non-circular Motions}
\label{s:dyn}

\cite{Trachternach08} report a measure of the relative non-circular motion in a subset of the THINGS galaxies; the total power in the non-circular harmonic components over the maximum velocity in the rotation curve ($A_r/v_{max}$). They report amplitudes of the non-circular motions lower than expected for the steep central mass density predicted by $\Lambda$CDM numerical simulations. This THINGS subset of 19 galaxies was selected to be non-interacting, limiting our comparison here. 
In Figure \ref{f:Ar}, we plot the morphological parameters as a function of relative non-circular strength. At best, there seem to be only very weak trends between Concentration, Asymmetry and $M_{20}$ with the relative non-circular motion ($A_r$). However, the highest value of $A_r$ (NGC 3627), is a close group member (there is no $A_r$ value for M51). Thus, morphological and dynamical parameters combined may delineate a superb space to identify mergers and harassed galaxies. Future large surveys will have the dynamical information automatically assessed by GALAPAGOS opening possibilities in combination with these morphological parameters.

\section{Discussion}
\label{s:disc}

In this paper we present our determinations of six morphological parameters computed within an inner and outer contour, approximately corresponding to the stellar and gas disk of these spirals. We aimed to determine how well \hi \ morphology can serve as an indicator of tidal interaction. There is much anecdotal evidence for this but the parameterization used to date mostly in restframe ultraviolet images offers a way to quantify. We selected the THINGS sample as it is nearby, mostly comprised of spirals and offers many excellent ancillary data. 
We tried to answer the following questions: which disk size (stellar or gas) works best to determine the \hi \ morphology in? How well does the \hi \ morphology compare to others used for interaction so far? Which parameters are the best indicators of interaction in \hi? 

\subsection{Which wavelength shows interaction?}

From the morphology-wavelength plots in the Appendix A ({\em online version of the paper}), as well as Figures \ref{f:fuv}, \ref{f:nuv}, and \ref{f:24}, it is apparent that the wavelength in which the parameter is determined is very important. Ultra-violet and the \hi \ parameters appear to correlate as these trace star-formation and some of its fuel or alternatively massive stars and the atomic component due to their photo-dissociation \citep{Allen04}.  Future \hi \ results on interaction rates of spirals should be very similar to ultra-violet determined ones. 

Indeed, the sensitivity of \hi \ observations to gas-rich and minor mergers are a critical benefit as these types dominate the mergers at higher redshift \citep{Lotz10a,Lotz10b}.
The all-sky surveys envisaged, will then provide a valuable local reference for the characterization of these types of mergers. The added benefits are that \hi \ appears to be very sensitive to gravitational interaction and there is no lag time waiting for star-formation to occur. On the other hand, \hi \ morphology may be uniquely sensitive to other phenomena, such as ram pressure stripping.

\subsection{Which disk size works best?}

The wavelength in which these parameters are determined is critical (see the Figures in Appendix A of individual galaxies ({\em online version of the paper}) or Figures \ref{f:fuv} -- \ref{f:24}), but which of the two choices to delineate the area for they are computed is optimal?
There is some extra information in optical morphological parameters if they are computed over the extent of the gas disk ($3 \times 10^{19}$ cm$^{-2}$ contour) as opposed to the stellar disk ($3 \times 10^{21}$ cm$^{-2}$ contour). Notably, ultra-violet, Spitzer 24 \mum, and especially of course the \hi \ change with increased area. In the case of UV, this is partly because the sky level and noise are so low that a signal is easier to pick out.  The compatibility is good news if we want to compare future \hi \ results, computed of the whole gas disk, to those found by other authors on restframe ultraviolet disks. 

\subsection{Which parameters are indicators of interaction?}

The main limitation to answer this question is the choice of sample as the THINGS sample was originally chosen as a relatively quiescent reference sample; most THINGS galaxies are group members. With the exception of M51, there are no ongoing mergers. In previous studies, both the Asymmetry and $M_{20}$ parameters were used to identify strongly perturbed systems in HST deep fields. The \hi \ value of these do seem to follow those in the FUV and NUV relatively closely, which is promising but not conclusive. If one compares the tidal parameter or our ranking for the bulk of the THINGS sample, there is a gentle rise in Asymmetry that one would not expect if this parameter is exclusively dependent on interaction yet the $M_{20}$ parameter for M51's stellar disk does not seem out of place with the rest of the THINGS and VIVA\footnote{VLA Imaging of Virgo spirals in Atomic gas, \citep{Chung09}.} sample. Interestingly, the $M_{20}$ value for the entire gas disk does seem to deviate somewhat (Figure \ref{f:theta}). Therefore, we speculate here that a combination of these two parameters may yield the best result in identifying merging or strongly interacting systems.
In our next papers, \cite{Holwerda10c} and \cite{Holwerda10d}, this speculation is borne out by a large sample of \hi \ maps and simulations of the \hi \ column density maps during a merger. 

\section{Conclusions}
\label{s:concl}

From our multi-wavelength analysis of the quantified morphology of \nsamp galaxies in the THINGS survey, we conclude the following: 

\begin{itemize}
\item[1.] The \hi \ morphology in column density maps down to approximately $10^{20}$ cm$^{-2}$ is promising perspective on galaxy interactions. Many of the \hi \ parameters show a close link to those in the near- and far-ultraviolet, which have been used to date to morphologically identify ongoing mergers. Concentration, Asymmetry and Ellipticity appear to be closely linked; Smoothness and $M_{20}$ show a noisier relation between \hi \ and UV morphological parameters and the Gini parameter is the notable exception with little correlation between the \hi \ and UV determined values. 
\item[2.] Despite the close link between far- and near-ultraviolet and \hi \ morphology, the wavelength in which the morphological parameter is measured is key, with the choice of area over which the parameters are computed of secondary importance. 
This has some implications for comparing morphological parameters across different studies, especially if these were done in different wavelengths. The only exception seems to be the UV and \hi \ morphologies, which can be compared more directly.
An increase in sensitivity of the \hi \ maps, or a corresponding decrease \hi \ contour level, only affects the Gini parameter due to the addition of many more low-value data points. The remaining morphological parameters are not affected. This is encouraging for studies comparing \hi \ morphologies in maps with different depths.
\item[3.] There are, at best, only weak trends between Concentration, Asymmetry, and $M_{20}$ and the non-circular motion strength ($A_r/v_{max}$) (Figure \ref{f:Ar}). This may imply that \hi \ morphological and kinematic deviations can be complementary tracers of disturbances of the spiral disk.
\item[4.] The two most common morphological parameters to identify mergers, Asymmetry and $M_{20}$, show contradictionary behaviour in \hi. Asymmetry seems to be very sensitive to disturbances, such as interaction, but also other phenomena (starburst etc.), while $M_{20}$ may be somewhat too insensitive (Figures \ref{f:theta} and \ref{f:rank}). 
Based on the THINGS sample, which admittedly does not span a sufficient range of gravitational interaction, we speculate that a combination of these two parameters may be useful for merger selection based on \hi \ morphology.
\end{itemize}

In the future, the high-resolution \hi \ maps of THINGS can serve as an excellent reference because these spirals are in just the type of small groups that the majority of spirals reside in. Bigger \hi \ surveys such as Westerbork \hi \ Survey Project  \citep[WHISP][]{whisp, whisp2}, and those to be undertaken with APERTIF, ASKAP, EVLA, MeerKAT, and eventually SKA will provide ever increasing statistics on the \hi \ morphology of spirals. Using some of these bigger samples, we hope to identify the part of \hi \ morphology space which correspond to ongoing mergers.
In the next papers in this series, we aim to quantify morphology to the WHISP sample on \hi \ column density maps, identify the merger space for \hi \ morphology, the timescale mergers are visible, and ultimately infer a merger rate based on the WHISP sample.

\section*{Acknowledgments}

We thank the THINGS, SINGS, SDSS and GALEX collaborations for making their data public. The authors would like to thank the anonymous referee for his or her careful work which led to a substantial improvement of this paper. 
The authors would like to thank W. Clarkson and J. Lotz for comments and discussions. 
We acknowledge support from the National Research Foundation of South Africa. The work of B.W. Holwerda  and W.J.G. de Blok is based upon research supported by the South African Research Chairs Initiative of the Department of Science and Technology and the National Research Foundation. A. Bouchard acknowledges the financial support from the South African Square Kilometre Array Project.
Research based on public data from GALEX, SDSS, Spitzer and the VLA.
The National Radio Astronomy Observatory is a facility of the National Science Foundation operated under cooperative agreement by Associated Universities, Inc. 
This work is based in part on observations made with the Spitzer Space Telescope, which is operated by the Jet Propulsion Laboratory, California Institute of Technology under a contract with NASA.
Funding for the SDSS and SDSS-II has been provided by the Alfred P. Sloan Foundation, the Participating Institutions, the National Science Foundation, the U.S. Department of Energy, the National Aeronautics and Space Administration, the Japanese Monbukagakusho, the Max Planck Society, and the Higher Education Funding Council for England. The SDSS Web Site is http://www.sdss.org/.
The SDSS is managed by the Astrophysical Research Consortium for the Participating Institutions. The Participating Institutions are the American Museum of Natural History, Astrophysical Institute Potsdam, University of Basel, University of Cambridge, Case Western Reserve University, University of Chicago, Drexel University, Fermilab, the Institute for Advanced Study, the Japan Participation Group, Johns Hopkins University, the Joint Institute for Nuclear Astrophysics, the Kavli Institute for Particle Astrophysics and Cosmology, the Korean Scientist Group, the Chinese Academy of Sciences (LAMOST), Los Alamos National Laboratory, the Max-Planck-Institute for Astronomy (MPIA), the Max-Planck-Institute for Astrophysics (MPA), New Mexico State University, Ohio State University, University of Pittsburgh, University of Portsmouth, Princeton University, the United States Naval Observatory, and the University of Washington. 
Based on observations made with the NASA Galaxy Evolution Explorer. GALEX is operated for NASA by the California Institute of Technology under NASA contract NAS5-98034.
This research has made use of the NASA/IPAC Extragalactic Database (NED) which is operated by the Jet Propulsion Laboratory, California Institute of Technology, under contract with the National Aeronautics and Space Administration.
In memory of Dr. M.J. Holwerda.


\begin{thebibliography}{81}
\expandafter\ifx\csname natexlab\endcsname\relax\def\natexlab#1{#1}\fi

\bibitem[{{Abraham} {et~al.}(1996{\natexlab{a}}){Abraham}, {Tanvir},
  {Santiago}, {Ellis}, {Glazebrook}, \& {van den Bergh}}]{Abraham96b}
{Abraham} R.~G., {Tanvir} N.~R., {Santiago} B.~X., {Ellis} R.~S., {Glazebrook}
  K., {van den Bergh} S., 1996{\natexlab{a}}, \mnras, 279, L47

\bibitem[{{Abraham} {et~al.}(1994){Abraham}, {Valdes}, {Yee}, \& {van den
  Bergh}}]{Abraham94}
{Abraham} R.~G., {Valdes} F., {Yee} H.~K.~C., {van den Bergh} S., 1994, \apj,
  432, 75

\bibitem[{{Abraham} {et~al.}(1996{\natexlab{b}}){Abraham}, {van den Bergh},
  {Glazebrook}, {Ellis}, {Santiago}, {Surma}, \& {Griffiths}}]{Abraham96a}
{Abraham} R.~G., {van den Bergh} S., {Glazebrook} K., {Ellis} R.~S., {Santiago}
  B.~X., {Surma} P., {Griffiths} R.~E., 1996{\natexlab{b}}, \apjs, 107, 1

\bibitem[{{Abraham} {et~al.}(2003){Abraham}, {van den Bergh}, \&
  {Nair}}]{Abraham03}
{Abraham} R.~G., {van den Bergh} S., {Nair} P., 2003, \apj, 588, 218

\bibitem[{{Allen} {et~al.}(2004){Allen}, {Heaton}, \& {Kaufman}}]{Allen04}
{Allen} R.~J., {Heaton} H.~I., {Kaufman} M.~J., 2004, \apj, 608, 314

\bibitem[{{Bendo} {et~al.}(2007){Bendo}, {Calzetti}, {Engelbracht},
  {Kennicutt}, {Meyer}, {Thornley}, {Walter}, {Dale}, {Li}, \&
  {Murphy}}]{Bendo07}
{Bendo} G.~J., {Calzetti} D., {Engelbracht} C.~W., {Kennicutt} R.~C., {Meyer}
  M.~J., {Thornley} M.~D., {Walter} F., {Dale} D.~A., {Li} A., {Murphy} E.~J.,
  2007, \mnras, 380, 1313

\bibitem[{{Bershady} {et~al.}(2000){Bershady}, {Jangren}, \&
  {Conselice}}]{Bershady00}
{Bershady} M.~A., {Jangren} A., {Conselice} C.~J., 2000, \aj, 119, 2645

\bibitem[{{Bertin} \& {Arnouts}(1996)}]{se}
{Bertin} E., {Arnouts} S., 1996, \aaps, 117, 393, provided by the NASA
  Astrophysics Data System

\bibitem[{{Boomsma} {et~al.}(2008){Boomsma}, {Oosterloo}, {Fraternali}, {van
  der Hulst}, \& {Sancisi}}]{Boomsma08}
{Boomsma} R., {Oosterloo} T.~A., {Fraternali} F., {van der Hulst} J.~M.,
  {Sancisi} R., 2008, \aap, 490, 555

\bibitem[{{Booth} {et~al.}(2009){Booth}, {de Blok}, {Jonas}, \&
  {Fanaroff}}]{MeerKAT}
{Booth} R.~S., {de Blok} W.~J.~G., {Jonas} J.~L., {Fanaroff} B., 2009, ArXiv
  e-prints/0910.2935

\bibitem[{{Bosma}(1981)}]{Bosma81b}
{Bosma} A., 1981, \aj, 86, 1825

\bibitem[{{Bournaud} {et~al.}(2005){Bournaud}, {Jog}, \&
  {Combes}}]{Bournaud05b}
{Bournaud} F., {Jog} C.~J., {Combes} F., 2005, \aap, 437, 69

\bibitem[{{Braun} {et~al.}(1994){Braun}, {Walterbos}, {Kennicutt}, \&
  {Tacconi}}]{Braun94}
{Braun} R., {Walterbos} R.~A.~M., {Kennicutt} Jr. R.~C., {Tacconi} L.~J., 1994,
  \apj, 420, 558

\bibitem[{{Bundy} {et~al.}(2005){Bundy}, {Ellis}, \& {Conselice}}]{Bundy05}
{Bundy} K., {Ellis} R.~S., {Conselice} C.~J., 2005, \apj, 625, 621

\bibitem[{{Carilli} \& {Rawlings}(2004)}]{ska}
{Carilli} C.~L., {Rawlings} S., 2004, New Astronomy Review, 48, 979

\bibitem[{{Chung} {et~al.}(2009){Chung}, {van Gorkom}, {Kenney}, {Crowl}, \&
  {Vollmer}}]{Chung09}
{Chung} A., {van Gorkom} J.~H., {Kenney} J.~D.~P., {Crowl} H., {Vollmer} B.,
  2009, ArXiv e-prints

\bibitem[{{Conselice}(2003)}]{CAS}
{Conselice} C.~J., 2003, \apjs, 147, 1

\bibitem[{{Conselice} {et~al.}(2000){Conselice}, {Bershady}, \&
  {Jangren}}]{Conselice00a}
{Conselice} C.~J., {Bershady} M.~A., {Jangren} A., 2000, \apj, 529, 886

\bibitem[{{Conselice} {et~al.}(2009){Conselice}, {Yang}, \&
  {Bluck}}]{Conselice09b}
{Conselice} C.~J., {Yang} C., {Bluck} A.~F.~L., 2009, \mnras, 361

\bibitem[{{Darling} \& {Giovanelli}(2006)}]{Darling06}
{Darling} J., {Giovanelli} R., 2006, \aj, 132, 2596

\bibitem[{{de Blok} {et~al.}(2009){de Blok}, {Jonas}, {Fanaroff}, {Holwerda},
  {Bouchard}, {Blyth}, {van der Heyden}, \& {Pirzkal}}]{meerkat2}
{de Blok} W.~J.~G., {Jonas} J., {Fanaroff} B., {Holwerda} B.~W., {Bouchard} A.,
  {Blyth} S., {van der Heyden} K., {Pirzkal} N., 2009, in Conference
  Proceedings of the "Panoramic Radio Astronomy: Wide-field 1-2 GHz research on
  galaxy evolution", June 02 - 05, 2009 Groningen, the Netherlands

\bibitem[{{de Blok} {et~al.}(2008){de Blok}, {Walter}, {Brinks},
  {Trachternach}, {Oh}, \& {Kennicutt}}]{de-Blok08}
{de Blok} W.~J.~G., {Walter} F., {Brinks} E., {Trachternach} C., {Oh} S.-H.,
  {Kennicutt} R.~C., 2008, \aj, 136, 2648

\bibitem[{{de Ravel} {et~al.}(2009){de Ravel}, {Le F{\`e}vre}, {Tresse},
  {Bottini}, {Garilli}, {Le Brun}, {Maccagni}, {Scaramella}, {Scodeggio},
  {Vettolani}, {Zanichelli}, {Adami}, {Arnouts}, {Bardelli}, {Bolzonella},
  {Cappi}, {Charlot}, {Ciliegi}, {Contini}, {Foucaud}, {Franzetti},
  {Gavignaud}, {Guzzo}, {Ilbert}, {Iovino}, {Lamareille}, {McCracken},
  {Marano}, {Marinoni}, {Mazure}, {Meneux}, {Merighi}, {Paltani}, {Pell{\`o}},
  {Pollo}, {Pozzetti}, {Radovich}, {Vergani}, {Zamorani}, {Zucca}, {Bondi},
  {Bongiorno}, {Brinchmann}, {Cucciati}, {de La Torre}, {Gregorini}, {Memeo},
  {Perez-Montero}, {Mellier}, {Merluzzi}, \& {Temporin}}]{de-Ravel09}
{de Ravel} L., {Le F{\`e}vre} O., {Tresse} L., {Bottini} D., {Garilli} B., {Le
  Brun} V., {Maccagni} D., {Scaramella} R., {Scodeggio} M., {Vettolani} G.,
  {Zanichelli} A., {Adami} C., {Arnouts} S., {Bardelli} S., {Bolzonella} M.,
  {Cappi} A., {Charlot} S., {Ciliegi} P., {Contini} T., {Foucaud} S.,
  {Franzetti} P., {Gavignaud} I., {Guzzo} L., {Ilbert} O., {Iovino} A.,
  {Lamareille} F., {McCracken} H.~J., {Marano} B., {Marinoni} C., {Mazure} A.,
  {Meneux} B., {Merighi} R., {Paltani} S., {Pell{\`o}} R., {Pollo} A.,
  {Pozzetti} L., {Radovich} M., {Vergani} D., {Zamorani} G., {Zucca} E.,
  {Bondi} M., {Bongiorno} A., {Brinchmann} J., {Cucciati} O., {de La Torre} S.,
  {Gregorini} L., {Memeo} P., {Perez-Montero} E., {Mellier} Y., {Merluzzi} P.,
  {Temporin} S., 2009, \aap, 498, 379

\bibitem[{{de Vaucouleurs} {et~al.}(1991){de Vaucouleurs}, {de Vaucouleurs},
  {Corwin}, {Buta}, {Paturel}, \& {Fouque}}]{RC3}
{de Vaucouleurs} G., {de Vaucouleurs} A., {Corwin} H.~G., {Buta} R.~J.,
  {Paturel} G., {Fouque} P., 1991, {Third Reference Catalogue of Bright
  Galaxies}. Volume 1-3, XII, 2069 pp.~7 figs..~ Springer-Verlag Berlin
  Heidelberg New York

\bibitem[{{Fraternali} {et~al.}(2002){Fraternali}, {van Moorsel}, {Sancisi}, \&
  {Oosterloo}}]{Fraternali02}
{Fraternali} F., {van Moorsel} G., {Sancisi} R., {Oosterloo} T., 2002, \aj,
  123, 3124

\bibitem[{{Frei} {et~al.}(1996){Frei}, {Guhathakurta}, {Gunn}, \&
  {Tyson}}]{Frei96}
{Frei} Z., {Guhathakurta} P., {Gunn} J.~E., {Tyson} J.~A., 1996, \aj, 111, 174

\bibitem[{{Giavalisco} {et~al.}(1996){Giavalisco}, {Livio}, {Bohlin},
  {Macchetto}, \& {Stecher}}]{Giavalisco96}
{Giavalisco} M., {Livio} M., {Bohlin} R.~C., {Macchetto} F.~D., {Stecher}
  T.~P., 1996, \aj, 112, 369

\bibitem[{Glasser(1962)}]{Glasser62}
Glasser G.~J., 1962, Journal of the American Statistical Association, 57, 648

\bibitem[{{Graham} \& {Driver}(2005)}]{Graham05}
{Graham} A.~W., {Driver} S.~P., 2005, Publications of the Astronomical Society
  of Australia, 22, 118

\bibitem[{{Guti{\'e}rrez} {et~al.}(2002){Guti{\'e}rrez},
  {L{\'o}pez-Corredoira}, {Prada}, \& {Eliche}}]{Gutierrez02}
{Guti{\'e}rrez} C.~M., {L{\'o}pez-Corredoira} M., {Prada} F., {Eliche} M.~C.,
  2002, \apj, 579, 592

\bibitem[{{Hibbard} {et~al.}(2001){Hibbard}, {van Gorkom}, {Rupen}, \&
  {Schiminovich}}]{Hibbard01}
{Hibbard} J.~E., {van Gorkom} J.~H., {Rupen} M.~P., {Schiminovich} D., 2001, in
  Astronomical Society of the Pacific Conference Series, Vol. 240, Gas and
  Galaxy Evolution, {Hibbard} J.~E., {Rupen} M., {van Gorkom} J.~H., eds., pp.
  657--+

\bibitem[{{Holwerda}(2005)}]{seman}
{Holwerda} B.~W., 2005, astro-ph/0512139

\bibitem[{{Holwerda} {et~al.}(2011{\natexlab{a}}){Holwerda}, {Pirzkal}, {de
  Blok}, {Blyth}, {Bouchard}, \& {van der Heyden}}]{Holwerda10e}
{Holwerda} B.~W., {Pirzkal} N., {de Blok} W.~J.~G., {Blyth} S.-L., {Bouchard}
  A., {van der Heyden} K.~J., 2011{\natexlab{a}}, \mnras, {\it submitted}

\bibitem[{{Holwerda} {et~al.}(2010){Holwerda}, {Pirzkal}, {de Blok}, {Blyth},
  {Bouchard}, {van der Heyden}, \& {Elson}}]{Holwerda10a}
{Holwerda} B.~W., {Pirzkal} N., {de Blok} W.~J.~G., {Blyth} S.-L., {Bouchard}
  A., {van der Heyden} K.~J., {Elson} E.~C., 2010, \mnras, {\it submitted}

\bibitem[{{Holwerda} {et~al.}(2011{\natexlab{b}}){Holwerda}, {Pirzkal}, {de
  Blok}, {Blyth}, {Bouchard}, {van der Heyden}, \& {Elson}}]{Holwerda10c}
---, 2011{\natexlab{b}}, \mnras, {\it accepted}

\bibitem[{{Holwerda} {et~al.}(2011{\natexlab{c}}){Holwerda}, {Pirzkal}, {de
  Blok}, {Blyth}, {Bouchard}, {van der Heyden}, \& {Elson}}]{Holwerda10d}
---, 2011{\natexlab{c}}, \mnras, {\it submitted}

\bibitem[{{Holwerda} {et~al.}(2011{\natexlab{d}}){Holwerda}, {Pirzkal}, {de
  Blok}, \& {van Driel}}]{Holwerda10f}
{Holwerda} B.~W., {Pirzkal} N., {de Blok} W.~J.~G., {van Driel} W.,
  2011{\natexlab{d}}, \mnras, {\it in preparation}

\bibitem[{{Hopkins} {et~al.}(2010){Hopkins}, {Croton}, {Bundy}, {Khochfar},
  {van den Bosch}, {Somerville}, {Wetzel}, {Keres}, {Hernquist}, {Stewart},
  {Younger}, {Genel}, \& {Ma}}]{Hopkins10}
{Hopkins} P.~F., {Croton} D., {Bundy} K., {Khochfar} S., {van den Bosch} F.,
  {Somerville} R.~S., {Wetzel} A., {Keres} D., {Hernquist} L., {Stewart} K.,
  {Younger} J.~D., {Genel} S., {Ma} C., 2010, ArXiv e-prints

\bibitem[{{Irwin} {et~al.}(2009){Irwin}, {Hoffman}, {Spekkens}, {Haynes},
  {Giovanelli}, {Linder}, {Catinella}, {Momjian}, {Koribalski}, {Davies},
  {Brinks}, {de Blok}, {Putman}, \& {van Driel}}]{Irwin09}
{Irwin} J.~A., {Hoffman} G.~L., {Spekkens} K., {Haynes} M.~P., {Giovanelli} R.,
  {Linder} S.~M., {Catinella} B., {Momjian} E., {Koribalski} B.~S., {Davies}
  J., {Brinks} E., {de Blok} W.~J.~G., {Putman} M.~E., {van Driel} W., 2009,
  \apj, 692, 1447

\bibitem[{{Jogee} {et~al.}(2009){Jogee}, {Miller}, {Penner}, {Skelton},
  {Conselice}, {Somerville}, {Bell}, {Zheng}, {Rix}, {Robaina}, {Barazza},
  {Barden}, {Borch}, {Beckwith}, {Caldwell}, {Peng}, {Heymans}, {McIntosh},
  {H{\"a}u{\ss}ler}, {Jahnke}, {Meisenheimer}, {Sanchez}, {Wisotzki}, {Wolf},
  \& {Papovich}}]{Jogee09}
{Jogee} S., {Miller} S.~H., {Penner} K., {Skelton} R.~E., {Conselice} C.~J.,
  {Somerville} R.~S., {Bell} E.~F., {Zheng} X.~Z., {Rix} H., {Robaina} A.~R.,
  {Barazza} F.~D., {Barden} M., {Borch} A., {Beckwith} S.~V.~W., {Caldwell}
  J.~A.~R., {Peng} C.~Y., {Heymans} C., {McIntosh} D.~H., {H{\"a}u{\ss}ler} B.,
  {Jahnke} K., {Meisenheimer} K., {Sanchez} S.~F., {Wisotzki} L., {Wolf} C.,
  {Papovich} C., 2009, \apj, 697, 1971

\bibitem[{{Johnston} {et~al.}(2008){Johnston}, {Taylor}, {Bailes}, {Bartel},
  {Baugh}, {Bietenholz}, {Blake}, {Braun}, {Brown}, {Chatterjee}, {Darling},
  {Deller}, {Dodson}, {Edwards}, {Ekers}, {Ellingsen}, {Feain}, {Gaensler},
  {Haverkorn}, {Hobbs}, {Hopkins}, {Jackson}, {James}, {Joncas}, {Kaspi},
  {Kilborn}, {Koribalski}, {Kothes}, {Landecker}, {Lenc}, {Lovell}, {Macquart},
  {Manchester}, {Matthews}, {McClure-Griffiths}, {Norris}, {Pen}, {Phillips},
  {Power}, {Protheroe}, {Sadler}, {Schmidt}, {Stairs}, {Staveley-Smith},
  {Stil}, {Tingay}, {Tzioumis}, {Walker}, {Wall}, \& {Wolleben}}]{askap}
{Johnston} S., {Taylor} R., {Bailes} M., {Bartel} N., {Baugh} C., {Bietenholz}
  M., {Blake} C., {Braun} R., {Brown} J., {Chatterjee} S., {Darling} J.,
  {Deller} A., {Dodson} R., {Edwards} P., {Ekers} R., {Ellingsen} S., {Feain}
  I., {Gaensler} B., {Haverkorn} M., {Hobbs} G., {Hopkins} A., {Jackson} C.,
  {James} C., {Joncas} G., {Kaspi} V., {Kilborn} V., {Koribalski} B., {Kothes}
  R., {Landecker} T., {Lenc} A., {Lovell} J., {Macquart} J.-P., {Manchester}
  R., {Matthews} D., {McClure-Griffiths} N., {Norris} R., {Pen} U.-L.,
  {Phillips} C., {Power} C., {Protheroe} R., {Sadler} E., {Schmidt} B.,
  {Stairs} I., {Staveley-Smith} L., {Stil} J., {Tingay} S., {Tzioumis} A.,
  {Walker} M., {Wall} J., {Wolleben} M., 2008, Experimental Astronomy, 22, 151

\bibitem[{{Jonas}(2007)}]{meerkat1}
{Jonas} J., 2007, in From Planets to Dark Energy: the Modern Radio Universe.
  October 1-5 2007, The University of Manchester, UK. Published online at
  SISSA, Proceedings of Science, p.7

\bibitem[{{Kamphuis} \& {Briggs}(1992)}]{Kamphuis92}
{Kamphuis} J., {Briggs} F., 1992, \aap, 253, 335

\bibitem[{{Karachentsev} {et~al.}(2004){Karachentsev}, {Karachentseva},
  {Huchtmeier}, \& {Makarov}}]{Karachentsev04}
{Karachentsev} I.~D., {Karachentseva} V.~E., {Huchtmeier} W.~K., {Makarov}
  D.~I., 2004, \aj, 127, 2031

\bibitem[{{Kennicutt} {et~al.}(2003){Kennicutt}, {Armus}, {Bendo}, {Calzetti},
  {Dale}, {Draine}, {Engelbracht}, {Gordon}, {Grauer}, {Helou}, {Hollenbach},
  {Jarrett}, {Kewley}, {Leitherer}, {Li}, {Malhotra}, {Regan}, {Rieke},
  {Rieke}, {Roussel}, {Smith}, {Thornley}, \& {Walter}}]{sings}
{Kennicutt} R.~C., {Armus} L., {Bendo} G., {Calzetti} D., {Dale} D.~A.,
  {Draine} B.~T., {Engelbracht} C.~W., {Gordon} K.~D., {Grauer} A.~D., {Helou}
  G., {Hollenbach} D.~J., {Jarrett} T.~H., {Kewley} L.~J., {Leitherer} C., {Li}
  A., {Malhotra} S., {Regan} M.~W., {Rieke} G.~H., {Rieke} M.~J., {Roussel} H.,
  {Smith} J.~T., {Thornley} M.~D., {Walter} F., 2003, \pasp, 115, 928

\bibitem[{{Kl{\"o}ckner} \& {Baan}(2005)}]{Klockner05}
{Kl{\"o}ckner} H., {Baan} W.~A., 2005, \apss, 295, 277

\bibitem[{{Lee} {et~al.}(2008){Lee}, {Kennicutt}, {Engelbracht}, {Calzetti},
  {Dale}, {Gordon}, {Dalcanton}, {Skillman}, {Begum}, {Funes}, {Gil de Paz},
  {Johnson}, {Sakai}, {van Zee}, {Walter}, {Weisz}, {Williams}, {Wu}, \&
  {Block}}]{lvl}
{Lee} J.~C., {Kennicutt} R.~C., {Engelbracht} C.~W., {Calzetti} D., {Dale}
  D.~A., {Gordon} K.~D., {Dalcanton} J.~J., {Skillman} E., {Begum} A., {Funes}
  J.~G., {Gil de Paz} A., {Johnson} B., {Sakai} S., {van Zee} L., {Walter} F.,
  {Weisz} D., {Williams} B., {Wu} Y., {Block} M., 2008, in Astronomical Society
  of the Pacific Conference Series, Vol. 396, Astronomical Society of the
  Pacific Conference Series, {Funes} J.~G., {Corsini} E.~M., eds., pp. 151--+

\bibitem[{{Lisker}(2008)}]{Lisker08}
{Lisker} T., 2008, ArXiv e-prints, 807

\bibitem[{{Lotz} {et~al.}(2008{\natexlab{a}}){Lotz}, {Davis}, {Faber},
  {Guhathakurta}, {Gwyn}, {Huang}, {Koo}, {Le Floc'h}, {Lin}, {Newman},
  {Noeske}, {Papovich}, {Willmer}, {Coil}, {Conselice}, {Cooper}, {Hopkins},
  {Metevier}, {Primack}, {Rieke}, \& {Weiner}}]{Lotz08b}
{Lotz} J.~M., {Davis} M., {Faber} S.~M., {Guhathakurta} P., {Gwyn} S., {Huang}
  J., {Koo} D.~C., {Le Floc'h} E., {Lin} L., {Newman} J., {Noeske} K.,
  {Papovich} C., {Willmer} C.~N.~A., {Coil} A., {Conselice} C.~J., {Cooper} M.,
  {Hopkins} A.~M., {Metevier} A., {Primack} J., {Rieke} G., {Weiner} B.~J.,
  2008{\natexlab{a}}, \apj, 672, 177

\bibitem[{{Lotz} {et~al.}(2008{\natexlab{b}}){Lotz}, {Jonsson}, {Cox}, \&
  {Primack}}]{Lotz08a}
{Lotz} J.~M., {Jonsson} P., {Cox} T.~J., {Primack} J.~R., 2008{\natexlab{b}},
  ArXiv e-prints, 805

\bibitem[{{Lotz} {et~al.}(2010{\natexlab{a}}){Lotz}, {Jonsson}, {Cox}, \&
  {Primack}}]{Lotz10a}
---, 2010{\natexlab{a}}, \mnras, 404, 590

\bibitem[{{Lotz} {et~al.}(2010{\natexlab{b}}){Lotz}, {Jonsson}, {Cox}, \&
  {Primack}}]{Lotz10b}
---, 2010{\natexlab{b}}, \mnras, 404, 575

\bibitem[{{Lotz} {et~al.}(2004){Lotz}, {Primack}, \& {Madau}}]{Lotz04}
{Lotz} J.~M., {Primack} J., {Madau} P., 2004, \aj, 128, 163

\bibitem[{{Malin} \& {Hadley}(1997)}]{Malin97b}
{Malin} D., {Hadley} B., 1997, in Astronomical Society of the Pacific
  Conference Series, Vol. 116, The Nature of Elliptical Galaxies; 2nd Stromlo
  Symposium, {Arnaboldi} M., {Da Costa} G.~S., {Saha} P., eds., pp. 460--+

\bibitem[{{Malin}(1978)}]{Malin78b}
{Malin} D.~F., 1978, \nat, 276, 591

\bibitem[{{Mu{\~n}oz-Mateos} {et~al.}(2009){Mu{\~n}oz-Mateos}, {Gil de Paz},
  {Zamorano}, {Boissier}, {Dale}, {P{\'e}rez-Gonz{\'a}lez}, {Gallego},
  {Madore}, {Bendo}, {Boselli}, {Buat}, {Calzetti}, {Moustakas}, \&
  {Kennicutt}}]{Munoz-Mateos09a}
{Mu{\~n}oz-Mateos} J.~C., {Gil de Paz} A., {Zamorano} J., {Boissier} S., {Dale}
  D.~A., {P{\'e}rez-Gonz{\'a}lez} P.~G., {Gallego} J., {Madore} B.~F., {Bendo}
  G., {Boselli} A., {Buat} V., {Calzetti} D., {Moustakas} J., {Kennicutt}
  R.~C., 2009, \apj, 703, 1569

\bibitem[{{Murphy} {et~al.}(2001){Murphy}, {Soifer}, {Matthews}, \&
  {Armus}}]{Murphy01}
{Murphy} Jr. T.~W., {Soifer} B.~T., {Matthews} K., {Armus} L., 2001, \apj, 559,
  201

\bibitem[{{Napier}(2006)}]{evla}
{Napier} P.~J., 2006, in Astronomical Society of the Pacific Conference Series,
  Vol. 356, Revealing the Molecular Universe: One Antenna is Never Enough,
  {D.~C.~Backer, J.~M.~Moran, \& J.~L.~Turner}, ed., pp. 65--+

\bibitem[{{Patton} {et~al.}(2000){Patton}, {Carlberg}, {Marzke}, {Pritchet},
  {da Costa}, \& {Pellegrini}}]{Patton00}
{Patton} D.~R., {Carlberg} R.~G., {Marzke} R.~O., {Pritchet} C.~J., {da Costa}
  L.~N., {Pellegrini} P.~S., 2000, \apj, 536, 153

\bibitem[{{Radburn-Smith} {et~al.}(2011){Radburn-Smith}, {de Jong}, {Seth},
  {Bell}, {Brown}, {Bullock}, {Courteau}, {Dalcanton}, {Ferguson},
  {Goudfrooij}, {Holfeltz}, {Holwerda}, {Purcell}, {Sick}, {Streich},
  {Vlaji{\'c}}, \& {Zucker}}]{ghosts}
{Radburn-Smith} D.~J., {de Jong} R.~S., {Seth} A.~C., {Bell} E.~F., {Brown}
  T.~M., {Bullock} J.~S., {Courteau} S., {Dalcanton} J.~J., {Ferguson} H.~C.,
  {Goudfrooij} P., {Holfeltz} S., {Holwerda} B.~W., {Purcell} C., {Sick} J.,
  {Streich} D., {Vlaji{\'c}} M., {Zucker} D.~B., 2011, \apj, {\it in
  preparation}

\bibitem[{{Ravindranath} {et~al.}(2006){Ravindranath}, {Giavalisco},
  {Ferguson}, {Conselice}, {Katz}, {Weinberg}, {Lotz}, {Dickinson}, {Fall},
  {Mobasher}, \& {Papovich}}]{Ravindranath06}
{Ravindranath} S., {Giavalisco} M., {Ferguson} H.~C., {Conselice} C., {Katz}
  N., {Weinberg} M., {Lotz} J., {Dickinson} M., {Fall} S.~M., {Mobasher} B.,
  {Papovich} C., 2006, \apj, 652, 963

\bibitem[{{Richter} \& {Sancisi}(1994)}]{Richter94}
{Richter} O.-G., {Sancisi} R., 1994, \aap, 290, L9

\bibitem[{{Sancisi} {et~al.}(2008){Sancisi}, {Fraternali}, {Oosterloo}, \& {van
  der Hulst}}]{Sancisi08}
{Sancisi} R., {Fraternali} F., {Oosterloo} T., {van der Hulst} T., 2008, \aapr,
  15, 189

\bibitem[{{Scarlata} {et~al.}(2007){Scarlata}, {Carollo}, {Lilly}, {Sargent},
  {Feldmann}, {Kampczyk}, {Porciani}, {Koekemoer}, {Scoville}, {Kneib},
  {Leauthaud}, {Massey}, {Rhodes}, {Tasca}, {Capak}, {Maier}, {McCracken},
  {Mobasher}, {Renzini}, {Taniguchi}, {Thompson}, {Sheth}, {Ajiki}, {Aussel},
  {Murayama}, {Sanders}, {Sasaki}, {Shioya}, \& {Takahashi}}]{Scarlata07}
{Scarlata} C., {Carollo} C.~M., {Lilly} S., {Sargent} M.~T., {Feldmann} R.,
  {Kampczyk} P., {Porciani} C., {Koekemoer} A., {Scoville} N., {Kneib} J.-P.,
  {Leauthaud} A., {Massey} R., {Rhodes} J., {Tasca} L., {Capak} P., {Maier} C.,
  {McCracken} H.~J., {Mobasher} B., {Renzini} A., {Taniguchi} Y., {Thompson}
  D., {Sheth} K., {Ajiki} M., {Aussel} H., {Murayama} T., {Sanders} D.~B.,
  {Sasaki} S., {Shioya} Y., {Takahashi} M., 2007, \apjs, 172, 406

\bibitem[{{SINGS team}(2006)}]{dr5}
{SINGS team}, 2006, Sings fifth data delivery release notes

\bibitem[{{Smith} {et~al.}(2007){Smith}, {Struck}, {Hancock}, {Appleton},
  {Charmandaris}, \& {Reach}}]{Smith07a}
{Smith} B.~J., {Struck} C., {Hancock} M., {Appleton} P.~N., {Charmandaris} V.,
  {Reach} W.~T., 2007, \aj, 133, 791

\bibitem[{{Springel} {et~al.}(2005){Springel}, {White}, {Jenkins}, {Frenk},
  {Yoshida}, {Gao}, {Navarro}, {Thacker}, {Croton}, {Helly}, {Peacock}, {Cole},
  {Thomas}, {Couchman}, {Evrard}, {Colberg}, \& {Pearce}}]{Springel05}
{Springel} V., {White} S.~D.~M., {Jenkins} A., {Frenk} C.~S., {Yoshida} N.,
  {Gao} L., {Navarro} J., {Thacker} R., {Croton} D., {Helly} J., {Peacock}
  J.~A., {Cole} S., {Thomas} P., {Couchman} H., {Evrard} A., {Colberg} J.,
  {Pearce} F., 2005, \nat, 435, 629

\bibitem[{{Stewart} {et~al.}(2000){Stewart}, {Fanelli}, {Byrd}, {Hill},
  {Westpfahl}, {Cheng}, {O'Connell}, {Roberts}, {Neff}, {Smith}, \&
  {Stecher}}]{Stewart00}
{Stewart} S.~G., {Fanelli} M.~N., {Byrd} G.~G., {Hill} J.~K., {Westpfahl}
  D.~J., {Cheng} K.-P., {O'Connell} R.~W., {Roberts} M.~S., {Neff} S.~G.,
  {Smith} A.~M., {Stecher} T.~P., 2000, \apj, 529, 201

\bibitem[{{Takamiya}(1999)}]{Takamiya99}
{Takamiya} M., 1999, \apjs, 122, 109

\bibitem[{{Taylor-Mager} {et~al.}(2007){Taylor-Mager}, {Conselice},
  {Windhorst}, \& {Jansen}}]{Taylor-Mager07}
{Taylor-Mager} V.~A., {Conselice} C.~J., {Windhorst} R.~A., {Jansen} R.~A.,
  2007, \apj, 659, 162

\bibitem[{{Trachternach} {et~al.}(2008){Trachternach}, {de Blok}, {Walter},
  {Brinks}, \& {Kennicutt}}]{Trachternach08}
{Trachternach} C., {de Blok} W.~J.~G., {Walter} F., {Brinks} E., {Kennicutt}
  R.~C., 2008, \aj, 136, 2720

\bibitem[{{Trujillo} {et~al.}(2001{\natexlab{a}}){Trujillo}, {Aguerri}, {Cepa},
  \& {Guti{\'e}rrez}}]{Trujillo01b}
{Trujillo} I., {Aguerri} J.~A.~L., {Cepa} J., {Guti{\'e}rrez} C.~M.,
  2001{\natexlab{a}}, \mnras, 328, 977

\bibitem[{{Trujillo} {et~al.}(2001{\natexlab{b}}){Trujillo}, {Aguerri},
  {Guti{\'e}rrez}, \& {Cepa}}]{Trujillo01}
{Trujillo} I., {Aguerri} J.~A.~L., {Guti{\'e}rrez} C.~M., {Cepa} J.,
  2001{\natexlab{b}}, \aj, 122, 38

\bibitem[{{Trujillo} {et~al.}(2007){Trujillo}, {Conselice}, {Bundy}, {Cooper},
  {Eisenhardt}, \& {Ellis}}]{Trujillo07}
{Trujillo} I., {Conselice} C.~J., {Bundy} K., {Cooper} M.~C., {Eisenhardt} P.,
  {Ellis} R.~S., 2007, \mnras, 382, 109

\bibitem[{{Trujillo} {et~al.}(2001{\natexlab{c}}){Trujillo}, {Graham}, \&
  {Caon}}]{Trujillo01a}
{Trujillo} I., {Graham} A.~W., {Caon} N., 2001{\natexlab{c}}, \mnras, 326, 869

\bibitem[{{van der Hulst}(2002)}]{whisp2}
{van der Hulst} J.~M., 2002, in Astronomical Society of the Pacific Conference
  Series, Vol. 276, Seeing Through the Dust: The Detection of HI and the
  Exploration of the ISM in Galaxies, {Taylor} A.~R., {Landecker} T.~L.,
  {Willis} A.~G., eds., pp. 84--+

\bibitem[{{van der Hulst} {et~al.}(2001){van der Hulst}, {van Albada}, \&
  {Sancisi}}]{whisp}
{van der Hulst} J.~M., {van Albada} T.~S., {Sancisi} R., 2001, in Astronomical
  Society of the Pacific Conference Series, Vol. 240, Gas and Galaxy Evolution,
  {Hibbard} J.~E., {Rupen} M., {van Gorkom} J.~H., eds., pp. 451--+

\bibitem[{{Verheijen} {et~al.}(2008){Verheijen}, {Oosterloo}, {van Cappellen},
  {Bakker}, {Ivashina}, \& {van der Hulst}}]{apertif}
{Verheijen} M.~A.~W., {Oosterloo} T.~A., {van Cappellen} W.~A., {Bakker} L.,
  {Ivashina} M.~V., {van der Hulst} J.~M., 2008, in American Institute of
  Physics Conference Series, Vol. 1035, The Evolution of Galaxies Through the
  Neutral Hydrogen Window, {R.~Minchin \& E.~Momjian}, ed., pp. 265--271

\bibitem[{{Walter} {et~al.}(2008){Walter}, {Brinks}, {de Blok}, {Bigiel},
  {Kennicutt}, {Thornley}, \& {Leroy}}]{Walter08}
{Walter} F., {Brinks} E., {de Blok} W.~J.~G., {Bigiel} F., {Kennicutt} R.~C.,
  {Thornley} M.~D., {Leroy} A., 2008, \aj, 136, 2563

\bibitem[{{Yan} {et~al.}(2005){Yan}, {Dickinson}, {Stern}, {Eisenhardt},
  {Chary}, {Giavalisco}, {Ferguson}, {Casertano}, {Conselice}, {Papovich},
  {Reach}, {Grogin}, {Moustakas}, \& {Ouchi}}]{Yan05}
{Yan} H., {Dickinson} M., {Stern} D., {Eisenhardt} P.~R.~M., {Chary} R.-R.,
  {Giavalisco} M., {Ferguson} H.~C., {Casertano} S., {Conselice} C.~J.,
  {Papovich} C., {Reach} W.~T., {Grogin} N., {Moustakas} L.~A., {Ouchi} M.,
  2005, \apj, 634, 109

\bibitem[{{Yitzhaki}(1991)}]{Yitzhaki91}
{Yitzhaki} S., 1991, American Statistical Association, 9, 235

\end{thebibliography}
\end{document}